\documentclass[journal]{IEEEtran}
\usepackage{amsmath,amsthm,amssymb,amsfonts,mathrsfs}
\usepackage{array}
\usepackage{textcomp}
\usepackage{stfloats}
\usepackage{url}
\usepackage{verbatim}
\usepackage{graphicx}
\usepackage{subfigure}
\usepackage{cite}
\usepackage{bm}
\usepackage{hyperref}

\usepackage{soul, color, xcolor}
\soulregister{\cite}7 
\soulregister{\citep}7 
\soulregister{\citet}7 
\soulregister{\ref}7 
\soulregister{\pageref}7 

\theoremstyle{definition}
\newtheorem{theorem}{\textbf{Theorem}}

\newtheorem{remark}{\textbf{Remark}}
\newtheorem{assumption}{\textbf{Assumption}}

\usepackage{threeparttable}
\usepackage{booktabs} 
\usepackage{multirow}

\usepackage[export]{adjustbox}

\usepackage{amsmath}
\usepackage{amssymb}

\usepackage[ruled,vlined,linesnumbered]{algorithm2e}

\begin{document}

\title{Differentiable Physics-Informed Adaptive Koopman Control for Stable Flight under Unknown Disturbances}

\author{Ao Jin$^{\dagger}$, 
	Qiujin Liang$^{\dagger}$,
	Tao Zhang, 
	Panfeng Huang, 
	Fan Zhang
	
	\thanks{$\dagger$ These authors contributed equally to this work. }
	\thanks{Ao Jin, Panfeng Huang and Fan Zhang are with Research Center for Intelligent Robotics, School of Astronautics, Northwestern Polytechnical University, Xi’an 710072, China. (E-mail: jinao@mail.nwpu.edu.cn)}
	\thanks{Qiujin Liang and Tao Zhang are with the Department of Automation, Tsinghua University, Beijing 100080, China.}
	}
\maketitle

\begin{abstract}
Uncertainties and disturbances in robotic systems, such as aerodynamic forces, are fundamentally outcomes of physical interactions with the environment, manifesting as learnable spatiotemporal sequences rather than random noise. However, achieving high-precision control for robotic systems operating in unstructured environments is often hindered by complex unmodeled dynamics and external disturbances. While learning-based methods offer powerful approximation capabilities, they typically suffer from heavy reliance on offline training and lack theoretical guarantees. Conversely, traditional robust control strategies are predominantly reactive, limited to instantaneous estimation without the foresight to anticipate future disturbance trends. To bridge this gap, this paper proposes a differentiable data-enabled Koopman control framework termed DEKC. Unlike black-box approaches, DEKC adopts a hybrid modeling strategy that retains the nominal physics model while employing a deep neural network to parameterize the lifting function of Koopman operator for unknown residual dynamics. Crucially, the framework formulates disturbances as a dynamical system, learning their temporal evolution in a global linear space. This enables the prediction of future disturbance trajectories, which are explicitly integrated into controller for preemptive compensation. Furthermore, an online backward gradient update mechanism is introduced to ensure real-time adaptation to time-varying uncertainties. Numerical simulations on a tethered space robot demonstrate the efficacy of the proposed DEKC in mitigating highly coupled uncertainties. Complementing these results, real-world experiments on a quadrotor substantiate its superiority in tracking agile trajectories under uncertainties induced by aerodynamics and suspended payload. 
\end{abstract}

\begin{IEEEkeywords}
Learning-Based Control, Disturbance Observer, Koopman Operator
\end{IEEEkeywords}

\section{Introduction}
Autonomous robotic systems, ranging from manipulators \cite{Deng2025a}, legged robots \cite{Grandia2023} and space robots to unmanned aerial vehicles \cite{OConnell2022,Chee2025,Sun2025}, are playing an increasingly pivotal role in real-world applications. A fundamental challenge spanning these diverse domains is the ability to operate robustly in unstructured environments characterized by complex interactions and significant uncertainties. Whether it is the nonlinear friction in robot joints, unknown terrain properties for legged locomotion, or external aerodynamic forces acting on aerial vehicles, accurate mathematical modeling of these uncertainties is often intractable in practice. Consequently, model-based controllers, which rely on precise prior knowledge of the system dynamics, frequently suffer from performance degradation or instability when facing such unmodeled dynamics.

Existing approaches to mitigate these uncertainties generally fall into two categories: model-based adaptive control and data-driven learning, yet both face significant limitations when applied to general agile robotic systems. Traditional robust control strategies, such as disturbance observers (DO) \cite{Pu2015,Chen2016,Mannini2025}, rely heavily on prior structural knowledge of the system dynamics. While effective for slowly varying or constant disturbances, these methods typically assume local linearity or bounded variation rates, which often fail to capture the highly nonlinear and fast-transient nature of complex interactions (e.g., aerodynamic turbulence or contact dynamics). Furthermore, they primarily focus on estimating the instantaneous disturbance for feedback compensation, lacking the capability to predict the future evolution of these disturbances, which is critical for proactive trajectory planning. Data-driven methods \cite{Torrente2021,OConnell2022,Qin2022,Chee2022,Richards2023a,Ferede2024,Das2025}, have been proven effective in handling arbitrary nonlinearities.  One of main benefits of these machine learning based methods is that they do not need any prior knowledge about the uncertainty and plant, but only the measured data of plant. This provides a more practical pattern of designing uncertainty observer for complicated tasks. However, their "black-box" nature presents severe challenges in terms of stability guarantees and sample efficiency. The high training cost and lack of interpretability make them risky for direct deployment on safety-critical physical hardware.

The uncertainties existed in real-world applications are always nonlinear and unknown. Koopman operator \cite{Koopman1931} is a tool that could map a nonlinear dynamical system to a linear system over an infinite dimensional Hilbert space. And the transformation can be implemented in a data-driven manner \cite{Proctor2018,Bevanda2021} with input/output data of system. This means that one can build an explicit and linear representation of unknown nonlinear system with measured data. Thus, the data-driven Koopman theory offers a new scheme to learn the unknown nonlinear dynamics in an explicit way \cite{Jia2023,Hao2024}. In this sense, the uncertainty in real-world applications which is unknown and nonlinear could be also captured with help of data-driven Koopman theory. 

However, the practical application of Koopman theory to general robotic systems is currently hindered by a critical bottleneck, the construction of an effective dictionary of observable functions. Standard approaches, such as Extended Dynamic Mode Decomposition (EDMD) \cite{Williams2015}, rely heavily on hand-crafted basis functions (e.g., polynomials, radial basis functions). Identifying the optimal set of basis functions for high-dimensional, strongly coupled robotic dynamics requires profound domain knowledge and extensive trial-and-error tunings. In addition, such fixed, user-defined bases often lack the sufficient representational ability to generalize across different scenario or to adapt to unstructured, time-varying disturbances (e.g., complex aerodynamic effects or shifting contact dynamics), thereby restricting the scalability and robustness of the method in real-world applications.

\begin{figure*}[htbp]
	\centering
	\includegraphics[width=36pc]{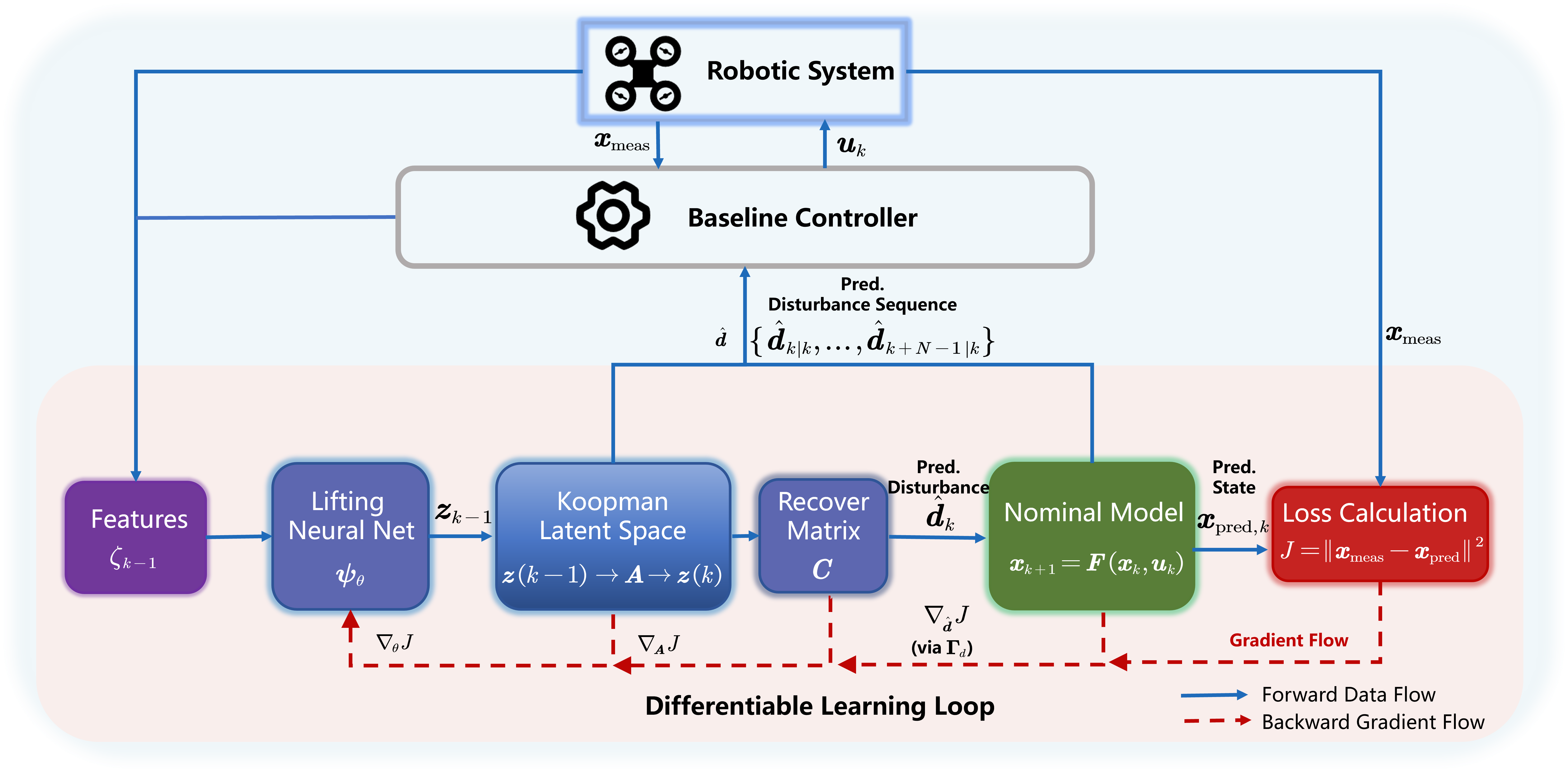} 
	\caption{The Differentiable Physics-Informed DEKC Framework. The architecture consists of a differentiable learning loop (blue region) and a control loop (green region). (a) Forward Pass: The lifting network $\bm{\psi}_\theta$ maps features to a latent space where dynamics are linear. The predicted disturbance $\bm{\hat{d}}$ is injected into the nominal physics model (differentiable simulator) to predict the future state $\bm{x}_{\text{pred}}$. (b) Differentiable Backward Pass (Red Dashed Lines): Unlike standard supervised learning, the gradients are not derived from disturbance labels but are backpropagated through the physics model (matrix $\bm{\Gamma}_d$). This "physics-aware" gradient $\nabla_{\bm{\hat{d}}} J$ updates the Koopman operator $\bm{A}$ and network parameters $\theta$ to minimize physical prediction errors directly, ensuring the learned disturbance is physically consistent with the system dynamics.} \label{framework}
\end{figure*}

In light of the considerations mentioned above, a differentiable data-enabled Koopman control framework called \textbf{\textit{DEKC}} for capturing uncertainties is proposed in this work. The focus of DEKC lies in formulating the unknown uncertainties as a linear dynamical system derived from Koopman operator theory and learning the temporal evolution directly from data. The proposed DEKC exploits the learned evolution to predict future trajectory of uncertainties. These predictions are then explicitly integrated into the controller for active compensation, thereby achieving automatic and preemptive suppression of complex coupled uncertainties. The predictive capability fundamentally distinguishes our approach from robust control methods such as DOBC, which primarily focus on estimating the instantaneous value of disturbances for reactive feedback.

Fig. \ref{framework} illustrated the proposed DEKC framework. The workflow consists of a forward inference path for robust control and a backward learning path for adaptation. The workflow of DEKC operates in a cycle of lifting, linear evolution, active compensation and backward gradient update. First, at time instant $k$, the feature vector is mapped by the lifting network $\bm{\psi_{\theta}}$ into a high-dimensional latent state $\bm{z}_k$. This step lifts the dynamics of uncertainty into a space where the evolution is globally linear. Second, the lifted state is propagated forward using the learned Koopman operator to generate a sequence of future latent states $\{\bm{z}_{k+1},\cdots,\bm{z}_{k+N}\}$. This linear prediction captures the latent trends of the uncertainty. Third, the recover matrix $\bm{C}$ reconstructs the predicted latent states back into the physical uncertainty domain $\{\hat{d}_{k+1}, \dots, \hat{d}_{k+N}\}$. These predictions can be injected into the feedback controller or model predictive controller (MPC) to preemptively offset the uncertainty. Last, to handle time-varying uncertainties, the framework executes a backward pass at each time step. A loss function ${J}$, penalizing the physical prediction errors, is computed. The gradients $\nabla_{\theta} \mathcal{L}$ and $\nabla_{A} \mathcal{L}$ are then backpropagated through the differentiable network structures to update the weights of the lifting network and the Koopman matrix in real-time. This ensures that the learned evolution continuously aligns with the changing physical environment. Notably, the DEKC does not employ fixed dictionary functions, but rather leverages a deep neural network to parameterize  a coordinate transformation that lifts the unknown nonlinear uncertainties into a linear space. This enables the extraction of optimal observable functions directly from data, effectively capturing the complex, coupled evolution of uncertainties (e.g., aerodynamic drag or suspended payload swing) without requiring prior structural assumptions.

\textbf{\textit{Main contributions of this work}}:
\begin{itemize}
	\item We challenge the conventional assumption that disturbances are instantaneous exogenous signals and instead model them as dynamical systems with predictable temporal evolution arising from robot–environment interactions.
	\item We propose a differentiable data-enabled Koopman control framework for capturing uncertainties, in which the Koopman operator and lifting function are updated online by backpropagating physical prediction errors through the nominal dynamics, ensuring consistency with system behavior under time-varying uncertainties.
	\item By capturing the temporal evolution of uncertainties in the lifted space, our method enables multi-step future prediction that is seamlessly integrated into both nonlinear feedback control and model predictive control, facilitating the proactive perception and compensation of future uncertainties.
	\item Through extensive numerical simulations and real-world quadrotor experiments involving aggressive maneuvers, aerodynamic effects, suspended payloads, and compound disturbances, we demonstrate that DEKC consistently outperforms state-of-the-art disturbance observers and learning-based estimators in accuracy, robustness, and real-time feasibility.
\end{itemize}

The rest of this paper is organized as follows. Section \ref{secion_preliminary} introduces the preliminaries and problem statement. Section \ref{section_learning_uncertainty} illustrates the method for learning uncertainty from data. Section \ref{section_augment_controller} presents the DEKC-augmented controller, followed by the stability analysis in Section \ref{section_stability}. Section \ref{section_simulation_A} and Section \ref{section_simulation_B} show the numerical evaluations. Section \ref{section_experiment} demonstrates the real-world validation results. Finally, Section \ref{section_conclusion} concludes this paper and discusses our future works.    

\section{Related Work}\label{section_related_works}
\subsection{Disturbance Estimation}
Accurate estimation and rejection of unmodeled dynamics and external uncertainties are fundamental for high-performance robotic control.

The proposition of disturbance observer based control (DOBC) \cite{Chen2016} offers a perspective for the control problem under consideration of uncertainty. Many disturbance observers (DO) are presented to estimate uncertainty, and compensate the output of it to the baseline controller such that the effect of disturbance can be attenuated or rejected. However, some of the existed disturbance observers \cite{Chen2004,Guo2005,Wei2010} assume that the disturbance can be represented by an exogenous system, and some prior information of the exogenous system is utilized in the design of disturbance observer. For the cases where the prior information of disturbance is unknown, these methods are no longer applicable. The literature \cite{Chen2000} proposes a disturbance observer that does not rely on the prior information of disturbance, but the disturbance is assumed to be varied slowly such that the first-order derivative of disturbance can be regarded as zero. This assumption is difficult to satisfy in practical scenarios, which limits its broader application. The extended state observer (ESO) \cite{Pu2015} treats the total disturbance as an augmented state variable and estimates it simultaneously with the system states. Similarly, adaptive control techniques \cite{Sastry1989} adjust controller gains or parameters online to accommodate parametric uncertainties.

To address complex, unstructured uncertainties, data-driven methods have been applied for estimating uncertainties. Gaussian Processes (GPs) \cite{Torrente2021} provide a non-parametric way to model residual dynamics with uncertainty bounds, often combined with MPC. Neural Networks (NNs) \cite{Bauersfeld2021,OConnell2022,Wei2025} have also been used to approximate the unknown inverse dynamics or residual terms directly. The literature \cite{Jia2023} considers the first-order derivative of disturbance in the disturbance observer, and learns the first-order derivative of disturbance by the Koopman theory. But the lifting functions of Koopman operator need to be selected before application and the observer gain requires to be re-designed across different tasks.

\subsection{Koopman Operator for Control}
The Koopman operator theory has emerged as a promising bridge between model-based control and data-driven control by globally linearizing nonlinear dynamics \cite{Shi2026}. This extends the application of standard linear control theory to the nonlinear systems, which significantly simplifies the controller design for nonlinear system. The works \cite{Korda2018a,Abraham2019,Alsalti2023,Mamakoukas2023,Eyuboglu2024} utilize data-driven Koopman theory for learning the dynamics of plant and use the learned dynamics for controller design. In these works, the dictionary of observables for the Koopman operator is selected manually, which requires deep domain expertise and is difficult to generalize across different robotic platforms.Recently, many learning-based methods have been proposed to identify finite-dimensional Koopman operators and to approximate the associated dictionary of observables. The works \cite{Lusch2018,Masti2021,Shi2022c,Bai2025,Singh2025} use the neural network to parameterize the dictionary of observables.

However, a majority of Koopman related works focus on learning the entire dynamics of the system from data. For well-understood robotic platforms like quadrotors or manipulators, discarding known physical laws in favor of a purely black-box Koopman model is data-inefficient and lacks interpretability. In contrast, our proposed DEKC framework adopts a hybrid modeling strategy. We retain the nominal physics model to handle the known dynamics and exclusively employ the deep Koopman operator to capture the unknown residual dynamics (disturbances). This decomposition significantly alleviates the learning burden and ensures that the controller preserves the reliability of first-principle models while gaining enhanced adaptability.

\subsection{Learning-Based Control}
To accommodate complex environmental interactions and unmodeled dynamics, learning-based control has emerged as a promising approach. Existing approaches can be broadly categorized into two classes: learning the system dynamics to enhance model-based control, and learning the control policy directly from data. 

The first class of learning-based control methods focuses on leveraging machine learning models to fit the system's transition function or residual dynamics, which are then integrated into a model-based controller \cite{Saviolo2022,Chee2022,Duong2024,Wei2025,Zhou2025}. 

The second class of learning-based control methods seeks to map state observations directly to control actions, bypassing the need for an explicit model. Reinforcement Learning (RL) have demonstrated impressive capabilities in solving complex tasks by maximizing long-term rewards \cite{Kaufmann2023,Eschmann2024,Jenelten2024,Zhang2025a}. Despite their success in simulation, RL policies are fundamentally ``black-box" systems, making it difficult to enforce strict state constraints or provide stability guarantees. Furthermore, the "Sim-to-Real" gap and low sample efficiency often hinder their direct deployment on physical hardware. To provide theoretical guarantees, recent works have integrated learning with rigorous control concepts. The works \cite{Sun2021,Rezazadeh2022,Singh2023} learn a contraction metric (parametrized by neural networks) to guarantee that all system trajectories converge exponentially to the nominal path. Similarly, neural Lyapunov methods \cite{Chang2019,Zinage2023} attempt to jointly learn a control policy and a Lyapunov function that certifies the stability of the closed-loop system. The fundamental limitation of these approaches lies in their heavy reliance on offline training with pre-collected datasets. Constructing a valid contraction metric or Lyapunov function typically requires an exhaustive exploration of the state space during the offline training phase.

\section{Preliminaries and Problem Statement} \label{secion_preliminary}
\subsection{Koopman Operator Theory: From Nonlinear to Linear}
Consider a discrete-time autonomous dynamical system evolving on a state space $\mathcal{X} \subseteq \mathbb{R}^n$
\begin{equation}
	\bm{x}_{k+1} = \bm{\mathcal{F}}(\bm{x}_k)
\end{equation}
where $\bm{x}_k$ is the state vector at time instant $k$ and $\bm{\mathcal{F}}$ is a nonlinear evolution map. Analyzing the global behavior of such nonlinear systems is often mathematically intractable. Koopman operator theory \cite{Koopman1931} offers an alternative perspective by lifting the dynamics from the state space to a space of observables. 

Let $\bm{g}: \mathcal{X} \to \mathbb{R}$ be a scalar-valued observable function belonging to an infinite-dimensional Hilbert space $\mathcal{H}$. The Koopman operator $\bm{\mathcal{K}}: \mathcal{H} \to \mathcal{H}$ is defined as the composition operator that advances the observable forward in time along the trajectory of the system
\begin{equation}
	\bm{\mathcal{K}} \bm{g}(\bm{x}_k) = \bm{g}(\bm{\mathcal{F}}(\bm{x}_k)) = \bm{g}(\bm{x}_{k+1})
\end{equation}
The most profound property of the Koopman operator is its linearity. Even though the underlying dynamics $\bm{\mathcal{F}}$ are nonlinear, the evolution of the observable $\bm{g}$ is governed by an infinite-dimensional linear map
\begin{equation}
	\bm{\mathcal{K}}(\alpha \bm{g}_1 + \beta \bm{g}_2) = \alpha \bm{\mathcal{F}} \bm{g}_1 + \beta \bm{\mathcal{F}} \bm{g}_2
\end{equation}
where $\alpha, \beta \in \mathbb{R}$ are arbitrary scalar constants, and $g_1, g_2$ are any observable functions lying in the lifting space. This linearity allows the application of powerful linear systems theory to analyze and control nonlinear systems globally.

\subsection{ Finite-Dimensional Approximation: From EDMD to Deep Learning}
Since the Koopman operator acts on an infinite-dimensional space, practical implementation requires projecting the dynamics onto a finite-dimensional invariant subspace $\mathcal{H}_N \subset \mathcal{H}$. Let $\bm{\Psi}(\bm{\zeta}) = [\psi_1(\bm{\zeta}), \dots, \psi_{N_L}(\bm{\zeta})]^\top$ denote a vector of basis functions (or lifting functions) that spans this subspace. The evolution is then approximated by a matrix $\bm{A} \in \mathbb{R}^{N_L \times N_L}$
\begin{equation}
	\bm{\Psi}(\bm{\zeta}_{k+1}) \approx \bm{A} \bm{\Psi}(\bm{\zeta}_k)
\end{equation}

Traditional approaches, such as extended dynamic mode decomposition (EDMD) \cite{Williams2015}, typically rely on a set of pre-defined dictionary of basis functions (e.g., polynomials, radial basis functions, or Fourier modes). While theoretically sound, EDMD faces two major challenges in complex robotic tasks: (1) The curse of dimensionality: The number of required basis functions grows exponentially with the system dimension; (2) Dependency on prior knowledge: The success of EDMD heavily depends on the manual selection of "good" observables that can capture the specific nonlinearity of the unknown dynamics.To overcome these limitations, the deep learning approach is adopted for learning the basis functions \cite{Lusch2018,Yeung2019,Mamakoukas2021,Xu2025}. Instead of fixing the dictionary $\bm{\Psi}(\cdot)$ a priori, it is parametrized by a deep neural network (DNN) denoted as $\bm{\Psi_}\theta(\cdot)$
\begin{equation}
	\bm{z}_k = \bm{\Psi}_\theta(\bm{\zeta}_k)
\end{equation}
where $\theta$ represents the trainable weights.

\subsection{Problem Formulation}
Consider a general nonlinear robotic system (e.g., quadrotors, manipulators, or tethered space robots) described by the following continuous-time control-affine dynamics
\begin{equation}
	\dot{\bm{x}}(t) = \bm{f}(\bm{x}(t)) + \bm{g}(\bm{x}(t))\bm{u}(t) + \bm{B}_d \bm{d} \label{system_with_disturbance}
\end{equation}
where $\bm{x}(t) \in \mathcal{X} \subseteq \mathbb{R}^n$ represents the state vector, $\bm{u}(t) \in \mathcal{U} \subseteq \mathbb{R}^m$ is the control input, and $\bm{d} \in \mathbb{R}^p$ represents the unknown lumped disturbance, which may include unmodeled dynamics, external forces (e.g., wind), or parametric uncertainties. The matrix $\bm{B}_d \in \mathbb{R}^{n \times p}$ is the known disturbance input matrix, characterizing how disturbances enter the system (matched or unmatched).

In practice, controllers are implemented in discrete time. We assume a sampling time $\Delta t$. The discretized dynamics can be expressed as
\begin{equation}
	\bm{x}_{k+1} = \bm{F}(\bm{x}_k, \bm{u}_k) + \bm{\Gamma}_d \bm{d}_k \label{eq_nominal_discrete_dynamics}
\end{equation}
where $\bm{F}(\cdot)$ represents the nominal dynamics (derived via Runge-Kutta or Euler integration of $\bm{f}(\bm{x})+\bm{g}(\bm{x})\bm{u}$), and $\bm{d}_k$ is the discrete-time equivalent of the disturbance. The matrix $\bm{\Gamma}_d \in \mathbb{R}^{n \times p}$ represents the discrete-time disturbance input matrix. It characterizes how the lumped disturbance $\bm{d}_k$ affects the state evolution during the sampling interval $\Delta t$. Depending on the discretization scheme employed (e.g., Forward Euler), $\bm{\Gamma}_d$ can be approximated as the continuous disturbance matrix scaled by the sampling time, i.e., $\bm{\Gamma}_d = \bm{B}_d \Delta t$.

\textbf{Problem Statement} Our goal is to design a disturbance predictor $\bm{\Phi}$ parameterized by $\theta$, which utilizes historical state and control data $\mathcal{H}_t = \{\bm{x}(\tau), \bm{u}(\tau) \mid \tau < t\}$ to generate a prediction $\hat{\bm{d}}(t)$
\begin{equation}
	\hat{\bm{d}}(t) = \bm{\Phi}( \mathcal{H}_t; \theta )
\end{equation}
such that the prediction error $\bm{e}_d(t) = \| \bm{d}(t) - \hat{\bm{d}}(t) \|$ is minimized asymptotically or bounded by a small constant $\epsilon$.

\section{Learning Uncertainty from Data}\label{section_learning_uncertainty}
This section elaborates the details of proposed learning-based scheme for capturing uncertainty from data.

\subsection{Modeling Disturbance Evolution via Koopman Operator}
While standard Koopman approaches focus on predicting the full system state $\bm{x}_k$, this work specifically targets the unknown external disturbance. Consider the control-affine system (\ref{system_with_disturbance}) subject to disturbances 
\begin{equation}
	\dot{\bm{x}}(t) = \bm{f}(\bm{x}(t)) + \bm{g}(\bm{x}(t))\bm{u}(t) + \bm{B}_d \bm{d} \notag
\end{equation}
Unlike standard robust control approaches that assume disturbances are bounded signals with no internal structure (i.e., $\|\bm{d}_k\|_\infty \le \delta$), we adopt a more structured perspective. We assume the disturbance is generated by an underlying, albeit unknown, dynamical process.

\begin{assumption}\label{ass_disturbance_dynamics}
	The disturbance $\bm{d}_k$ is not stochastic white noise, but is generated by an underlying (unknown) dynamical process (e.g., aerodynamic drag, wind gusts, or unmodeled interactions). This implies the existence of a latent evolution
	\begin{equation}
		\begin{cases}
			\bm{\zeta}_{k+1} = \bm{\mathcal{T}}(\bm{\zeta}_k) \\
			\quad \bm{d}_k = \bm{h}(\bm{\zeta}_k)
		\end{cases}
	\end{equation}
	where $\bm{\zeta}_k$ represents the augmented features relevant to the disturbance (e.g., state, control, and past disturbance values), and $\bm{\mathcal{T}}$ is the unknown nonlinear transition map.
\end{assumption}

By applying Koopman operator theory to this specific disturbance sub-dynamics, we seek a set of vector-valued observables $\bm{\Psi}(\bm{\zeta}) = [\psi_1(\bm{\zeta}), \dots, \psi_{N_L}(\bm{\zeta})]^\top$. In the space spanned by these observables, the nonlinear evolution of the disturbance features becomes linear
\begin{equation}
	\bm{\Psi}(\bm{\zeta}_{k+1}) = \bm{\mathcal{K}} \bm{\Psi}(\bm{\zeta}_k) 
\end{equation}
Consequently, the disturbance $\bm{d}_k$ can be reconstructed as a linear combination of these observables. This transforms the difficult problem of identifying the nonlinear map $\bm{\mathcal{T}}$ into a linear prediction problem in the observable space.

\begin{remark}
	The Assumption \ref{ass_disturbance_dynamics}, in which disturbances are generated by an underlying dynamical system, is physically well-grounded for most robotic systems. Unlike sensor noise which may approximate white noise, systemic disturbances—such as aerodynamic drag, friction, or suspended payload oscillations—possess inertia and temporal correlation. For instance, wind gusts evolve according to fluid dynamics, and unmodeled structural vibrations follow mass-spring-damper laws. These phenomena exhibit colored noise characteristics with distinct spectral patterns rather than flat-spectrum white noise. Leveraging this latent evolution ($\bm{\zeta}_{k+1} = \bm{\mathcal{T}}(\bm{\zeta}_k)$), we are inspired to design a learnable framework specifically tailored to capture and exploit the inherent predictability of these physical interactions.
\end{remark}

\subsection{Deep Neural Approximation of Koopman Embedding}
The core of the proposed framework is to discover a coordinate transformation (embedding function) $\bm{\psi}: \mathbb{R}^{n_x + n_u + n_d} \to \mathbb{R}^{N_L}$ that maps the augmented state-action-disturbance space into a higher-dimensional latent space where the disturbance evolution is linear. Since the analytical form of this transformation is unknown and highly nonlinear, we approximate it using a deep neural network, denoted as $\bm{\psi}_\theta$, where $\theta$ represents the network weights. 

To ensure physical interpretability, the lifting function $\bm{\psi}_{\theta}$ is constructed by explicitly concatenating the physical disturbance $\bm{d}_k$ with the learned auxiliary observables. Formally, the lifting process is given by
\begin{equation}
	\bm{z}_k = \bm{\psi}_{\theta}(\bm{\zeta}_k) = \begin{bmatrix} \bm{d}_k \\ \bm{\varphi}_{\text{mlp}}(\bm{\zeta}_k) \end{bmatrix} \in \mathbb{R}^{N_L}
	\label{eq_lifting_structure}
\end{equation}
where $\bm{d}_k \in \mathbb{R}^{n_d}$ is the disturbance component extracted from the input features $\bm{\zeta}_k$, and $\bm{\varphi}_{\text{mlp}}: \mathbb{R}^{n_{\zeta}} \to \mathbb{R}^{N_L - n_d}$ represents the deep neural network. Its architecture consists of an input layer, multiple hidden layers, and a linear output layer. The layer normalization is applied to the input feature vector $\bm{\zeta}_k$ to stabilize the training dynamics and ensure robustness against varying scales of state and disturbance inputs. The network contains hidden layers with nonlinear activation functions. We select the hyperbolic tangent function ($\tanh(\cdot)$) as the activation function because its bounded output range and smoothness are conducive to learning stable dynamical behaviors compared to ReLU. The final layer maps the features to the lifted space $\mathbb{R}^{N_L}$. The neural network  $\bm{\varphi}_{\text{mlp}}$ can be written as
\begin{equation}
	\bm{\varphi}_{\text{mlp}}(\bm{\zeta}_k) = W_{\text{out}} \cdot \sigma(\dots \sigma(LN(\bm{\bm{\zeta}}_k))\dots) + b_{\text{out}}
\end{equation}
where $\sigma(\cdot)$ is the activation function, and $LN(\cdot)$ denotes layer normalization.

\subsection{Data-Enable Disturbance Observation in the Lifted Space}
In the lifted space, the dynamics of unknown disturbance are governed by the Koopman operator $\bm{\mathcal{K}}$, parameterized as a learnable matrix $\bm{A} \in \mathbb{R}^{N_L \times N_L}$. The evolution is strictly linear
\begin{equation}
	\bm{z}_{k+1|k} = \bm{A} \bm{z}_k \label{eq_lls}
\end{equation}
The physical disturbance is then recovered via a linear output matrix $\bm{C} \in \mathbb{R}^{n_d \times N_L}$
\begin{equation}
	\hat{\bm{d}}_{k+1|k} = \bm{C} \bm{z}_{k+1|k} \label{eq_lls_recover}
\end{equation}
This structure allows us to predict the future disturbance sequence $\{\hat{\bm{d}}_{k+i|k}\}_{i=1}^{N}$ by simply powering the matrix $\bm{A}$, i.e., $\hat{\bm{d}}_{k+i|k} = \bm{C} \bm{A}^i \bm{\psi}_\theta(\bm{\zeta}_k)$, which is computationally efficient. In addition, the lifting function (\ref{eq_lifting_structure}) is structured such that the first $n_d$ states correspond to the physical disturbance. Consequently, the output matrix $\boldsymbol{C}$ is defined as a projection matrix
\begin{equation}
	\boldsymbol{C} = \begin{bmatrix} \boldsymbol{I}_{n_d} & \mathbf{0}_{n_d \times (N_L - n_d)} \end{bmatrix}
\end{equation}
where $\boldsymbol{I}_{n_d}$ denotes the identity matrix of dimension $n_d$, and $\mathbf{0}$ represents a zero matrix of compatible dimensions.

With the future disturbance sequence, the predicted state propagated from the nominal dynamics (\ref{eq_nominal_discrete_dynamics}) can be written by
\begin{equation}
	\hat{\bm{x}}_{k+1}=\bm{F}(\bm{x}_k,\bm{u}_k)+\bm{\Gamma}_d \hat{\bm{d}}_{k|k-1} \label{eq_predicted_dekc}
\end{equation}
where $\hat{\bm{d}}_{k|k-1} = \bm{C} \bm{A}_{k-1} \bm{z}_{k-1}$ is the disturbance predicted from the previous step. Unlike offline learning methods that require large datasets, our framework adapts the Koopman operator and the lifting network online to capture time-varying disturbances. We derive the explicit update laws based on the instantaneous physical prediction error
\begin{equation}
	\bm{e}_k = \bm{x}_{\text{meas}, k} - \hat{\bm{x}}_{k}
\end{equation}
which is defined as discrepancy between the measured state $\bm{x}_{\text{meas}, k}$ and the predicted state $\hat{\bm{x}}_{k}$ propagated from the nominal dynamics with the DEKC prediction.

We seek to update the parameters of Koopman operator and lifting network to minimize the instantaneous quadratic loss
\begin{equation}
	J_k = \frac{1}{2} \| \bm{e}_k \|_2^2 \label{loss_J}
\end{equation}
Based on the chain rule, the gradient of the loss $J$ with respect to the matrix $\bm{A}$ is computed by
\begin{equation}
	\frac{\partial J}{\partial \bm{A}} = \frac{\partial J}{\partial \hat{\bm{x}}_{k+1}} \cdot \frac{\partial \hat{\bm{x}}_{k+1}}{\partial \hat{\bm{d}}_k} \cdot \frac{\partial \hat{\bm{d}}_k}{\partial \hat{\bm{z}}_k} \cdot \frac{\partial \hat{\bm{z}}_k}{\partial \bm{A}}
\end{equation}
The sensitivity of the error to the disturbance prediction is 
\begin{equation}
	\frac{\partial J}{\partial \hat{\bm{d}}} = -\bm{\Gamma}_d \bm{e}_k
\end{equation}
Propagating this backwards through the linear output layer $\bm{C}$ and the Koopman evolution by combing (\ref{eq_lls}), (\ref{eq_lls_recover}) and (\ref{eq_predicted_dekc}), we obtain the update rule for matrix $A$
\begin{equation}
	\bm{A}_k = \bm{A}_{k-1} + \eta_A \frac{\mathcal{\bm{E}}_k \bm{z}_{k-1}^\top}{\|\bm{z}_{k-1}\|_2^2 + \epsilon} \label{update_A}
\end{equation}
where $\eta_A$ is the learning rate, $\epsilon$ is a small regularization term, and $\mathcal{\bm{E}}_k$ is the backpropagated error signal in the lifted space
\begin{equation}
	\mathcal{\bm{E}}_k = \bm{C}^\top \bm{\Gamma}_d^\top \bm{e}_k
\end{equation}
This update law resembles a normalized Least Mean Squares (LMS) filter, effectively projecting the physical error into the Koopman operator space to correct the linear evolution dynamics.

And the gradient of the loss $J$ with respect to the parameters $\theta$ of lifting function is computed by
\begin{equation}
	\frac{\partial J}{\partial \theta} = \frac{\partial J}{\partial \hat{\bm{x}}_{k+1}}\cdot \frac{\partial \hat{\bm{x}}_{k+1}}{\partial \hat{\bm{d}}_k} \cdot \frac{\partial \hat{\bm{d}}_k}{\partial \hat{\bm{z}}_k}\cdot \frac{\partial \hat{\bm{z}}_k}{\partial \bm{z}_{k-1}}\cdot \frac{\partial \bm{z}_{k-1}}{\partial \theta}
\end{equation}
To adapt the embedding manifold, we backpropagate the error signal through the current Koopman matrix. The error gradient with respect to the lifted state $\bm{z}_{k-1}$ is given by 
\begin{equation}
	\frac{\partial J}{\partial \bm{z}_{k-1}} = -\bm{A}_{k-1}^\top \mathcal{\bm{E}}_k
\end{equation}
Now, the parameters $\theta$ of the lifting network are updated using stochastic gradient descent (SGD)
\begin{equation}
	\theta_k = \theta_{k-1} + \eta_\theta \left( \nabla_\theta \bm{\psi}( \bm{\zeta}_{k-1} ) \right)^\top \bm{A}_{k-1}^\top C^\top \bm{\Gamma}_d^\top \bm{e}_k \label{update_theta}
\end{equation}
where the term $\nabla_\theta \bm{\psi}( \bm{\zeta}_{k-1} )$ can be obtained using automatic differentiation tools (e.g., PyTorch, Jax). This explicit backpropagation ensures that the learned embedding $\bm{\psi}_\theta$ evolves to linearize the disturbance dynamics that are most relevant to the physical system's dynamic characteristic. 

To prevent the learned dynamics from diverging, we impose a hard constraint on the spectral norm of $\bm{A}$. After each update, if $\|\bm{A}\|_F > \gamma$, we rescale the matrix
\begin{equation}
	\bm{A}_k \leftarrow \gamma \frac{\bm{A}_k}{\|\bm{A}_k\|_F}
\end{equation}
This ensures the bounded-input bounded-output (BIBO) stability of the disturbance predictor.

By jointly learning the network parameters $\theta$ and the transition matrix $\bm{A}$, the proposed framework automatically discovers an optimal, low-dimensional coordinate system where the disturbance dynamics appear linear.

\section{DEKC-Augmented Controller}\label{section_augment_controller}
The proposed DEKC framework operates as a universal, data-driven disturbance observer. Its "Plug-and-Play" architecture allows it to function independently of the specific controller structure, providing real-time disturbance estimates $\hat{d}$ (and future trajectories) that can be seamlessly injected into various control schemes to recover nominal performance. To demonstrate this versatility, we integrate DEKC with two distinct control paradigms: (1) Nonlinear Feedback Control, representing high-frequency explicit control laws suitable for computationally constrained platforms; and (2) Model Predictive Control, which computes the optimal control action by solving a finite-horizon constrained optimization problem based on a predictive model of the system. 

\subsection{DEKC-Augmented Nonlinear Feedback Control}
To robustify the nominal controller, we propose a composite control law that explicitly cancels the unknown disturbance.
\begin{assumption}
	The disturbance $\bm{d}$ satisfies the matching condition, and there exists a nominal feedback control law $\bm{u}_{\text{nom}}=\bm{\pi}(\bm{x})$\footnote{The baseline controller $\bm{\pi}(\bm{x})$ can be derived by using feedback linearization, backstepping, or geometric control on Lie groups.} that renders the nominal system (where $\bm{d}=\bm{0}$) asymptotically stable. 
\end{assumption}

As DEKC operates in discrete time with sampling interval $\Delta t$, we use a zero-order hold (ZOH) assumption for the inter-sample disturbance prediction $\hat{\bm{d}}_k$. Now, the DEKC-Feedback law is formulated as
\begin{equation}
	\bm{u}(t) = \bm{\pi}(\bm{x}(t))\underbrace{- \bm{g}^{\dagger}(\bm{x}(t)) \bm{B}_d \hat{\bm{d}}_k}_{\text{Feedforward Compensation}}, \quad t \in [t_k, t_{k+1}) \label{dekc_feedback_controller}
\end{equation}
where $\bm{g}^{\dagger}(\bm{x})$ denotes the Moore-Penrose pseudoinverse of the input matrix. 

By injecting this compensation term, the closed-loop dynamics approximate the ideal nominal system. The effective disturbance seen by the feedback controller is reduced from the full magnitude $\|d\|$ to the DEKC prediction error $\|\tilde{\bm{d}}\| = \|\bm{d} - \hat{\bm{d}}_k\|$, significantly enhancing tracking precision.

\subsection{DEKC-Augmented Model Predictive Control}
For scenarios requiring explicit constraint handling and foresight, we integrate DEKC into an MPC framework. The detailed formulation is stated as follows.

At time step $k$, DEKC lifts the current feature vector $\zeta_k$ and evolves it linearly to generate a sequence of future disturbance predictions over the horizon $N$
\begin{equation}
	\mathcal{D}_{\text{pred}} = \{ \hat{\bm{d}}_{k|k}, \hat{\bm{d}}_{k+1|k}, \dots, \hat{\bm{d}}_{k+N-1|k} \} \label{pred_disturbance}
\end{equation}
where $\hat{\bm{d}}_{k+i|k} = \bm{C} \bm{A}^i \bm{\psi}_\theta(\bm{\zeta}_k)$. The sequence is (\ref{pred_disturbance}) injected into the MPC formulation as a time-varying external parameter. The Optimal Control Problem (OCP) is defined as
\begin{equation} \label{mpc_dekc}
	\begin{aligned}
		\min_{\bm{U}_k} \quad & \sum_{i=0}^{N-1} L(\bm{x}_{k+i|k}, \bm{u}_{k+i|k}) + V_f(\bm{x}_{k+N|k}) \\
		\text{s.t.} \quad & \bm{x}_{k|k} = \bm{x}_k \\
		& \bm{x}_{k+i+1|k} = \bm{F}(\bm{x}_{k+i|k}, \bm{u}_{k+i|k}) + \bm{\Gamma}_d \hat{\bm{d}}_{k+i|k} \\
		& \bm{u}_{\text{min}} \leq \bm{u}_{k+i|k} \leq \bm{u}_{\text{max}} \\
		& \bm{x}_{\text{min}} \leq \bm{x}_{k+i|k} \leq \bm{x}_{\text{max}}
	\end{aligned}
\end{equation}
where $L$ is the stage cost, $V_f$ is the terminal cost. The nominal discrete dynamics $\bm{F}(\cdot)$ are corrected by the predicted disturbance sequence. This enables the MPC to optimize control inputs that proactively counteract the predicted disturbance profile (e.g., wind gusts), effectively neutralizing their impact on the system dynamics.

\begin{algorithm}[t]
	\caption{DEKC-Augmented Model Predictive Control}
	\label{alg:dekc}
	\SetAlgoLined
	\DontPrintSemicolon
	
	\KwIn{Nominal dynamics $\bm{x}_{k+1} = \bm{f}_{\text{nom}}(\bm{x}_k, \bm{u}_k)$,  parameters of lifting function $\bm{\psi}$ and Koopman matrix $\bm{A}$, recover matrix $\bm{C}$, learning rate $\eta$, horizon $N$.}
	\KwOut{Augmented control input $\bm{u}^*_k$.}
	
	\For{time step $k = 0, 1, 2, \dots$}{
		Measure current system state $x_k$.\;
		Calculate historical residual disturbance (mismatch):\;
		\Indp $\bm{d}_k = \bm{x}_k - \bm{f}_{\text{nom}}(\bm{x}_{k-1}, \bm{u}_{k-1})$\;
		\Indm
		
		Lift disturbance to latent space: $\bm{z}_{k|k} = \bm{\psi}_{\theta}(\bm{\zeta}_k)$\;
		Generate predicted disturbance sequence for horizon $N$:\;
		\Indp
		\For{$i = 0$ \KwTo $N-1$}{
			Evolve in latent space: $\bm{z}_{k+i+1|k} = \bm{A} \bm{z}_{k+i|k}$.\;
			Recover to physical space: $\hat{d}_{k+i+1|k} = \bm{C}\bm{z}_{k+i+1|k}$.\;
		}
		\Indm
		
		Solve the optimization problem (\ref{mpc_dekc}) to obtain $\bm{u}^*_k$.\;
		
		Apply control input $\bm{u}^*_k$ to the robot.\;
		
		Compute physical prediction error (\ref{loss_J}).\;
		Update parameters of lifting function and Koopman matrix via (\ref{update_theta}) and (\ref{update_A}).   \;
	}
\end{algorithm}

\section{Stability Analysis}\label{section_stability}
In this section, we provide a unified stability analysis for both control schemes. We prove that under a bounded estimation error from the DEKC framework, both the DEKC-Feedback and DEKC-MPC schemes are input-to-state stable (ISS).

\subsection{Stability of DEKC-Feedback System}
\begin{assumption} \label{dekc_estimation_error_bounded}
	(Bounded Estimation Error): The DEKC estimation error $\tilde{\bm{d}}(t) = \bm{d}(t) - \hat{\bm{d}}_k$ for $t\in [t_k,t_{k+1})$ is uniformly bounded, i.e., $\|\tilde{\bm{d}}(t)\| \le \epsilon$ for all $t \ge 0$.
\end{assumption}

\begin{theorem}
	Consider the nonlinear system (\ref{system_with_disturbance}) controlled by the DEKC-Feedback law (\ref{dekc_feedback_controller}). If the nominal controller $\bm{\pi}(\bm{x})$ renders the disturbance-free system exponentially stable, then the closed-loop system is ISS with respect to the estimation error $\epsilon$.
\end{theorem}
\begin{proof}
	Let $V(\bm{x})$ be a Lyapunov function for the nominal closed-loop system satisfying
	\begin{equation}
		\begin{aligned}
			\alpha_1(\|\bm{x}\|) \le V(\bm{x}) \le \alpha_2(\|\bm{x}\|)\\
			\quad \frac{\partial V}{\partial \bm{x}} (\bm{f}(\bm{x}) + \bm{g}(\bm{x})\bm{\pi}(\bm{x})) \le -\alpha_3(\|\bm{x}\|)
		\end{aligned}
	\end{equation}
	where $\alpha_i$ are class $\mathcal{K}_\infty$ functions. The time derivative of $V$ along the trajectories of the actual system (\ref{system_with_disturbance}) with control law (\ref{dekc_feedback_controller}) is
	\begin{equation}
		\begin{aligned}
			\dot{V} &= \frac{\partial V}{\partial \bm{x}} \left( \bm{f}(\bm{x}) + \bm{g}(\bm{x}) \left[ \bm{\pi}(\bm{x}) - \bm{g}^{\dagger}(\bm{x}) \bm{B}_d \hat{\bm{d}} \right] + \bm{B}_d \bm{d} \right) \\
			&= \frac{\partial V}{\partial \bm{x}} \left( \bm{f}(\bm{x}) + \bm{g}(\bm{x})\bm{\pi}(\bm{x}) \right) + \frac{\partial V}{\partial \bm{x}} \left( \bm{B}_d \bm{d} - \bm{g}(x)\bm{g}^{\dagger}(\bm{x})\bm{ B}_d \hat{\bm{d}} \right)
		\end{aligned}
	\end{equation}
	Under the matching condition, $\bm{g} \bm{g}^{\dagger} \bm{B}_d = \bm{B}_d$. Thus
	\begin{equation}
		\dot{V} = \frac{\partial V}{\partial x} \left( \bm{f}(\bm{x}) + \bm{g}(\bm{x})\bm{\pi}(\bm{x}) \right) + \frac{\partial V}{\partial \bm{x}} \bm{B}_d (\bm{d} - \hat{\bm{d}})
	\end{equation}
	Using the bounds on $V$ and the Cauchy-Schwarz inequality
	\begin{equation}
		\dot{V} \le -\alpha_3(\|\bm{x}\|) + \left\| \frac{\partial V}{\partial \bm{x}} \bm{B}_d \right\| \|\tilde{\bm{d}}\| \le -\alpha_3(\|\bm{x}\|) + L_V \|\bm{B}_d\| \epsilon
	\end{equation}
	where $L_V$ is the Lipschitz constant of $V$. The system state $\bm{x}$ will converge to the set $\Omega = \{ \bm{x} \mid \alpha_3(\|\bm{x}\|) \le L_V \|\bm{B}_d\| \epsilon \}$. As the DEKC network minimizes $\epsilon$ via online learning, the ultimate bound $\Omega$ shrinks towards the origin.
\end{proof}

\subsection{Stability of DEKC-MPC System}
In this subsection, we establish the ISS properties of the closed-loop system under the proposed DEKC-MPC scheme. Since the DEKC estimator provides an approximation of the true disturbance, the estimation error acts as a bounded perturbation to the nominal system. We formally prove that the system state remains bounded within a compact set proportional to this estimation error.

Let the true discrete-time dynamics of the system be
\begin{equation}
	\bm{x}_{k+1} = \bm{F}(\bm{x}_k, \bm{u}_k) + \bm{\Gamma}_d \bm{d}_k \label{dekc_mpc_nominal}
\end{equation}
The DEKC-MPC relies on a prediction model that incorporates the estimated disturbance $\hat{\bm{d}}_k$
\begin{equation}
	\hat{\bm{x}}_{k+1} = \bm{F}(\bm{x}_k, \bm{u}_k) + \bm{\Gamma}_d \hat{\bm{d}}_k \label{dekc_mpc_dekc}
\end{equation}
Define the disturbance estimation error as $\bm{e}_k \triangleq \bm{d}_k - \hat{\bm{d}}_k$. By subtracting (\ref{dekc_mpc_dekc}) from (\ref{dekc_mpc_nominal}), the true state at $k+1$ relates to the predicted state by
\begin{equation}
	\bm{x}_{k+1} = \hat{\bm{x}}_{k+1} + \bm{\Gamma}_d \bm{e}_k
\end{equation}

To proceed with the stability proof, we adopt standard assumptions from robust MPC theory \cite{Rawlings2017}.

\begin{assumption} \label{ass_nominal_stability_cost}
	(Nominal Stability Cost) Let $V_N(\bm{x}_k)$ be the optimal value function of the nominal MPC optimization problem at time $k$. The stage cost $L(\bm{x},\bm{ u}) = \|\bm{x}\|_Q^2 + \|\bm{u}\|_R^2$ and the terminal cost $V_f(\bm{x})$ are chosen such that $V_N(\cdot)$ serves as a Lyapunov function for the nominal system. Specifically, there exist $\mathcal{K}_\infty$-functions $\alpha_1, \alpha_2, \alpha_3$ such that for all $\bm{x} \in \mathcal{X}$
	\begin{equation}
		\begin{aligned}
			&\alpha_1(\|\bm{x}\|) \le V_N(\bm{x}) \le \alpha_2(\|\bm{x}\|) \\
			&V_N(\bm{F}(\bm{x}, \bm{u}^*(\bm{x})) + \bm{\Gamma}_d \hat{\bm{d}}) - V_N(\bm{x}) \le -\alpha_3(\|\bm{x}\|)
		\end{aligned} 
	\end{equation}
	where $\bm{u}^*(\bm{x})$ is the optimal feedback control law derived from the MPC.
\end{assumption}

\begin{assumption} \label{lipschitz_value_function}
	(Lipschitz Continuity of Value Function) The optimal value function $V_N(\bm{x})$ is locally Lipschitz continuous in the domain of interest $\mathcal{X}$. That is, there exists a Lipschitz constant $L_V > 0$ such that for any $\bm{x}, \bm{y} \in \mathcal{X}$
	\begin{equation}
		|V_N(\bm{x}) - V_N(\bm{y})| \le L_V \|\bm{x} - \bm{y}\| \label{ineq_lipschitz}
	\end{equation}
\end{assumption}

\begin{theorem}
	Under Assumptions \ref{dekc_estimation_error_bounded}-\ref{lipschitz_value_function}, the closed-loop system controlled by the DEKC-MPC is ISS  with respect to the disturbance estimation error $e_k$. The system state $\bm{x}_k$ converges to a robust positively invariant set $\Omega$ defined by the error bound $\epsilon$.
\end{theorem}

\begin{proof}
	We examine the evolution of the Lyapunov function candidate $V_N(\bm{x})$ along the trajectories of the true system (\ref{dekc_mpc_nominal}). Consider the difference in the value function from time $k$ to $k+1$
	\begin{equation}
		\Delta V = V_N(\bm{x}_{k+1}) - V_N(\bm{x}_k)
	\end{equation}
	Adding and subtracting the value function of the optimal predicted state $\hat{\bm{x}}_{k+1}^* = \bm{F}(\bm{x}_k, \bm{u}_k^*) + \bm{\Gamma}_d \hat{\bm{d}}_k$, we get
	\begin{equation}
		\Delta V = \underbrace{V_N(\bm{x}_{k+1}) - V_N(\hat{\bm{x}}_{k+1}^*)}_{\text{Term 1: Perturbation Effect}} + \underbrace{V_N(\hat{\bm{x}}_{k+1}^*) - V_N(\bm{x}_k)}_{\text{Term 2: Nominal Descent}}  \label{eq_Delta_V}
	\end{equation}
	For the Term 2 (Nominal Descent), from Assumption \ref{ass_nominal_stability_cost}, the MPC controller ensures the nominal system energy decreases
	\begin{equation}
		V_N(\hat{\bm{x}}_{k+1}^*) - V_N(\bm{x}_k) \le -\alpha_3(\|\bm{x}_k\|) \label{ineq_nominal_descent}
	\end{equation}
	For the Term 1 (Perturbation Effect), using Assumption \ref{lipschitz_value_function} and the relationship in (\ref{ineq_lipschitz}), we can derive
	\begin{equation}
		\begin{aligned}
			&V_N(\bm{x}_{k+1}) - V_N(\hat{\bm{x}}_{k+1}^*) \le L_V \| \bm{x}_{k+1} - \hat{\bm{x}}_{k+1}^* \| \\
			&= L_V \| (\bm{F}(\bm{x}_k, \bm{u}_k^*) + \bm{\Gamma}_d \bm{d}_k) - (\bm{F}(\bm{x}_k, \bm{u}_k^*) + \bm{\Gamma}_d \hat{\bm{d}}_k) \| \\
			&= L_V \| \bm{\Gamma}_d (\bm{d}_k - \hat{\bm{d}}_k) \| \\
			&= L_V \| \bm{\Gamma}_d \bm{e}_k \| \\
			&\le L_V \| \bm{\Gamma}_d \| \epsilon
		\end{aligned} \label{ineq_perturbation_effect}
	\end{equation}
	Substituting (\ref{ineq_nominal_descent}) and (\ref{ineq_perturbation_effect}) into (\ref{eq_Delta_V}), we obtain the ISS inequality
	\begin{equation}
		V_N(\bm{x}_{k+1}) - V_N(\bm{x}_k) \le -\alpha_3(\|\bm{x}_k\|) + L_V \|\bm{\Gamma}_d\| \epsilon  \label{ineq_Delta_V}
	\end{equation}
	From the above inequality, we observe that the Lyapunov function decreases (i.e., $\Delta V < 0$) as long as
	\begin{equation}
		\alpha_3(\|\bm{x}_k\|) > L_V \|\bm{\Gamma}_d\| \epsilon  \label{ineq_mpc_stable_condition}
	\end{equation}
	Let $\upsilon = \alpha_3^{-1}(L_V \|\bm{\Gamma}_d\| \epsilon)$. Equation (\ref{ineq_mpc_stable_condition}) implies that whenever $\|\bm{x}_k\| > \upsilon $, the system state will be driven towards the origin. Consequently, the state $\bm{x}_k$ ultimately converges to the bounded set
	\begin{equation}
		\Omega = \{ \bm{x} \in \mathcal{X} \mid \|\bm{x}\| \le \alpha_3^{-1}(L_V \|\bm{\Gamma}_d\| \epsilon) \}
	\end{equation}
	This completes the proof. 
\end{proof}

\begin{remark}
	The size of the ultimate bound $\Omega$ is directly proportional to the estimation error $\epsilon$. The proposed DEKC framework minimizes $\epsilon$ through two mechanisms: (1) Koopman Embedding: The linear evolution in the lifted space $\bm{z}_{k+1} = \bm{A} \bm{z}_k$ captures the global disturbance dynamics more effectively; (2) Online Adaptation: The explicit parameter update laws (derived in Section \ref{section_learning_uncertainty}-D) continuously minimize the instantaneous error $\bm{e}_k$, ensuring that $\epsilon$ remains small even under time-varying disturbance conditions. Therefore, compared to standard robust MPC where $\epsilon$ is determined by the worst-case disturbance bound $\sup \|\bm{d}_k\|$, our method reduces the bound to the worst-case prediction error, resulting in significantly tighter control accuracy.
\end{remark}

\section{Numerical Evaluation-A}\label{section_simulation_A}
In this section, we select the tethered space robot (TSR) as the numerical validation example, as it represents a critical class of space applications with significant potential for missions such as space-debris removal, artificial-gravity generation, and asteroid exploration \cite{Cartmell2008, Huang2018a}. Beyond its practical significance, the TSR serves as an ideal testbed for evaluating the proposed framework due to the highly coupled nonlinear dynamics between the platform and the TSR, combined with the complex oscillatory disturbances induced by the flexible tether. To rigorously evaluate the proposed framework, we benchmark it against three representative disturbance observers: the adaptive disturbance observer \cite{Wang2017a}, the nonlinear disturbance observer \cite{Li2014a} and the sliding-mode disturbance observer \cite{Shtessel2007,Lu2012}. The primary motivation for this comparison is to highlight a fundamental limitation of these conventional methods. They are predominantly reactive, relying on instantaneous error feedback to estimate the current disturbance. In contrast, for systems like TSR where disturbances exhibit distinct temporal patterns (e.g., tether vibrations), the ability to learn and predict the disturbance evolution is crucial. This comparative analysis aims to demonstrate how the predictive nature of DEKC translates into superior disturbance suppression and tracking accuracy compared to the passive adaptation of baseline observers.

\subsection{Dynamics of TSR}
As \cite{Williams2008} shows, the in-plane dynamics of tethered space robot are highly nonlinear and its mathematical model is described by
\begin{equation}
	\begin{aligned}\label{tsr_dynamics}
		{\alpha}''+2\frac{l'}{l}(\Omega+\alpha ') + 3\Omega^2\cos \alpha \sin\alpha = 0 \\
		l'' - l[(\Omega+\alpha')^2+(3\cos^2 \theta -1)\Omega^2] = -\frac{T}{{m}} 
	\end{aligned}
\end{equation}
where $m$ is the mass of tethered space robot, $\alpha$ is the in-plane angle, $\Omega$ is the orbit angular velocity, $l$ and $T$ are the instantaneous length and the tension of space tether respectively, $()'$  denotes the derivation with respect to the time $t$. By introducing the transformation from time $t$ to the true anomaly $\tau$, i.e., $\tau=\Omega t$, the dimensionless form of (\ref{tsr_dynamics}) can be written as
\begin{equation}
	\begin{aligned}
		\ddot{\alpha}+2\frac{\dot{\lambda}}{\lambda+1}(1+\dot{\alpha})+3\cos \alpha \sin\alpha = 0
		\\
		\ddot{\lambda} - (\lambda+1) [(1+\dot{\alpha})^2-1+3\cos^2\alpha] = -u
	\end{aligned}
\end{equation}
and
\begin{equation}
	\lambda = l/L-1,\ \tau = \Omega t, \ u =  T / (\Omega^2 L {m})
\end{equation}
where $\lambda\in (-1,0]$ is the tether deployment ratio, $u\ge 0$ is the control input,  $\tau$ is the dimensionless form of time $t$, $L$ is the total length of space tether, $\dot{()}$ represents the derivation with respect to the dimensionless time $\tau$. Furthermore, if we consider the unknown disturbance $\bm d$ (such as oscillations in tether tension induced by tether vibrations), the dynamics can be written as a general form
\begin{equation}
	\dot{\bm{x}} = \bm{f}(\bm x) + \bm B_u \bm u +  \bm B_d \bm d \label{tsr_state_space_dynamics}
\end{equation}
where $\bm x = [x_1,x_2,x_3,x_4]$, $\bm u=[u]$, $\bm x\in \mathcal{X}:=\{\bm x \in \mathbb{R}^n|-1<y_2\le0\}$, $\bm u \in \mathcal{U}:=\{\bm u \in \mathbb{R}^p|u \ge0\}$, $n=4$, $p=1$, $\bm d \in \mathcal{W}$ and $\mathcal{W}$ is compact. And $\bm f(\bm x)$ and $\bm B_u$ are defined by
\begin{gather} 
	\bm{f}(\bm{x})=
	\left[ \begin{array}{c}
		x_3\\
		x_4\\
		-2\frac{x_4}{x_2+1}(x_3+1)-3\sin x_1\cos x_1\\
		\left( x_2+1 \right) \left[ \left( x_3+1 \right) ^2+3\cos ^2x_1-1 \right]\\
	\end{array} \right] \notag
	\\
	\ \ \ \ \ \ \bm{B}_u=\bm{B}_d=\begin{bmatrix}
		0 & 0 & 0 & -1
	\end{bmatrix}^\mathrm{T}
	\label{dynamics_dimensionless}
\end{gather}
where $x_1 = \alpha$, $x_2 =\lambda$, $x_3 = \dot{\alpha}$, $x_4 = \dot{\lambda}$.

\begin{figure}[!t]
	\centering
	{\includegraphics[width=13pc]{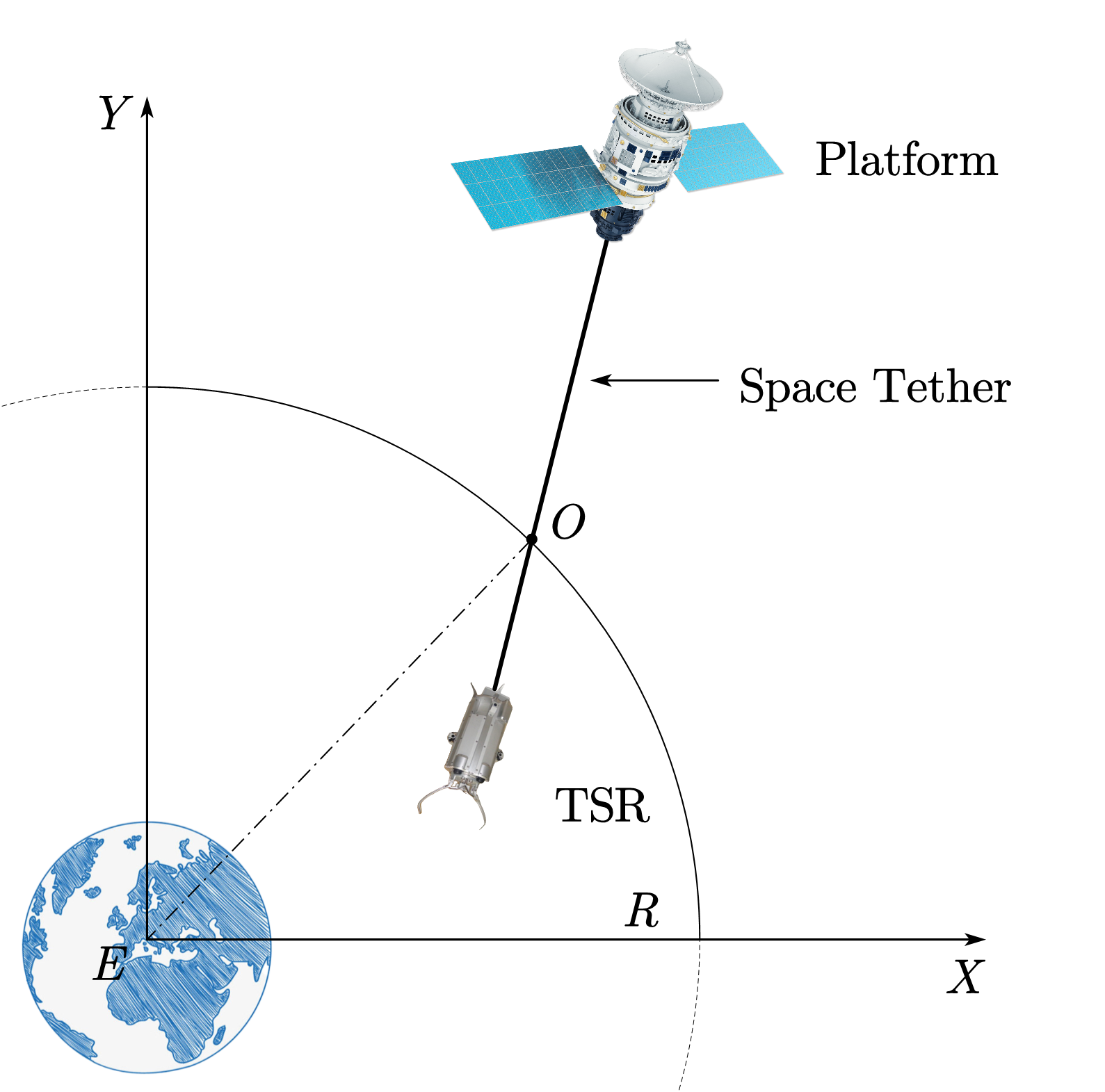}}
	\caption{The tethered space robot is connected to the platform by a space tether.}
	\label{tsr}
\end{figure}

\subsection{Implementation Details}
\textit{Simulation Setup}: The shared representations (lifting functions) are defined by $\bm{\psi}=[\bm{d}\ \bm{\phi}(\bm{\zeta})]$, where $\bm{\phi}$ is parameterized by a two-layer deep neural network with a hidden layer comprising 64 neurons. The activation function of deep neural network is selected as hyperbolic tangent function (Tanh) \footnote{$a(x) = \tanh(x)$}. The dimension of output of $\bm{\psi}$ is set to 24, i.e. $N=24$. The feature vector is set to $\bm{\zeta}=[\bm{d}\ \bm{x}\ \bm{u}]$. The recover matrix is $\bm{C}=[\bm{I}_p\ \bm{0}_{K-p}]$. The learning rate is $0.05$. All training in this work is implemented with Pytorch \cite{Paszke2019}. And the dynamics of TSR is discretized by using Runge-Kutta four method with sampling interval $t_s=0.01\ \text{s}$. The total simulation time is $T = 10\ \text{s}$. The feedback controller for deploying TSR is designed by \cite{Vadali1991a}
\begin{equation}
	\begin{aligned} \label{tsr_nominal_controller}
		\bm{\pi}(\bm{x}) &= 3(x_2+1)+k_1x_2 \\
		&-\frac{2}{3}k_2 x_3\frac{1+x_3}{1+x_2}+ k_3 x_4
	\end{aligned}
\end{equation}
where $k_1=1$, $k_2=0.5$, and $k_3=3.5$. The initial state of TSR is set to  $[0\ 0\ -0.99\ 1.5]^\mathrm{T}$.

\textit{Disturbance Scenarios}: Three types of external disturbances with increasing complexity are considered
\begin{align}
	\textit{A}(t) &= 0.6(\cos(3t)+\sin(2t)), \label{d1}\\[4pt]
	\textit{B}(\bm{x},t) &= 0.5(\cos(12t)+\sin(4t)+\sin(4x_2)\label{d2} \\&+\cos(x_4)+\sin(x_4))\nonumber, \\[4pt]
	\textit{C}(\bm{x},t) &= \textit{B}(\bm{x},t) + 0.8 H(t - t_s), \label{d3}
\end{align}
where the function $H(\cdot)$ denotes the Heaviside step function with step time $t_s=8$ s, introducing an abrupt change in the third disturbance scenario. These three cases correspond to:   
\begin{itemize}
	\item \textit{A(t)}: a purely time-varying sinusoidal disturbance;
	\item \textit{B(x,t)}: a disturbance that depends on both the system states and time;
	\item \textit{C(x,t)}: a state- and time-dependent disturbance containing an additional step component.
\end{itemize}
These three cases are designed to represent increasing levels of complexity and nonlinearity in disturbance characteristics. 

\textit{Baseline Disturbance Observers}: To assess the effectiveness of the proposed algorithm, we compare it with three widely used baseline disturbance observers: an adaptive disturbance observer (ADO) \cite{Wang2017a}, a nonlinear disturbance observer (NDO) \cite{Li2014a} and a sliding-mode disturbance observer (SMDO) \cite{Shtessel2007,Lu2012}. Each observer is tested under all three disturbance scenarios to ensure a fair and comprehensive comparison. And the parameters of the ADO, NDO, and SMDO are individually tuned to achieve their best performance on the TSR system under identical conditions. 

\begin{table*}[t]
	\centering
	\caption{Parameter settings of the baseline observers and the proposed method} \label{baseline_do_params}
	\begin{tabular}{lccc}
		\toprule
		\textbf{Observer} & \textbf{Parameter} & \textbf{Symbol} & \textbf{Value} \\
		\midrule
		
		\multirow{2}{*}{Adaptive Disturbance Observer (ADO)}
		& Adaptation Gain & $a$     & $12$ \\
		& Sampling Period &$T_s$    & $0.01$ \\
		
		\midrule
		
		\multirow{1}{*}{Nonlinear Disturbance Observer (NDO)}
		& Observer Gain & $L$    & $diag [10.0\ 10.0\ 10.0\ 10.0]$ \\
		
		\midrule
		
		\multirow{2}{*}{Sliding Mode Disturbance Observer (SMDO)}
		& Sliding Gain  &$L$      & $diag[0.3\ 0.1\ 0,1\ 0.1]$ \\
		& Observer Gain   &$\lambda$ 	  & $diag[30.0\ 15.0\ 1.4]$ \\
		
		\midrule
		
		\multirow{2}{*}{Proposed Method}
		& Learning Rate     & $\eta$     & $0.05$ \\
		& Dimension of Lifting Function & $K$  & $24$ \\
		
		\bottomrule
	\end{tabular}
\end{table*}

\textit{Evaluation Metrics}: We evaluate the performance of ADO, NDO, SMDO, and the proposed algorithm in terms of disturbance estimation accuracy $\text{RMSE}(\hat{\bm{d}},\bm{d})$, stabilization error $||\bm{x}||^2$ and convergence characteristics.

\subsection{Evaluation Results}
\begin{figure}[!t]
	\centering
	{\includegraphics[width=21pc]{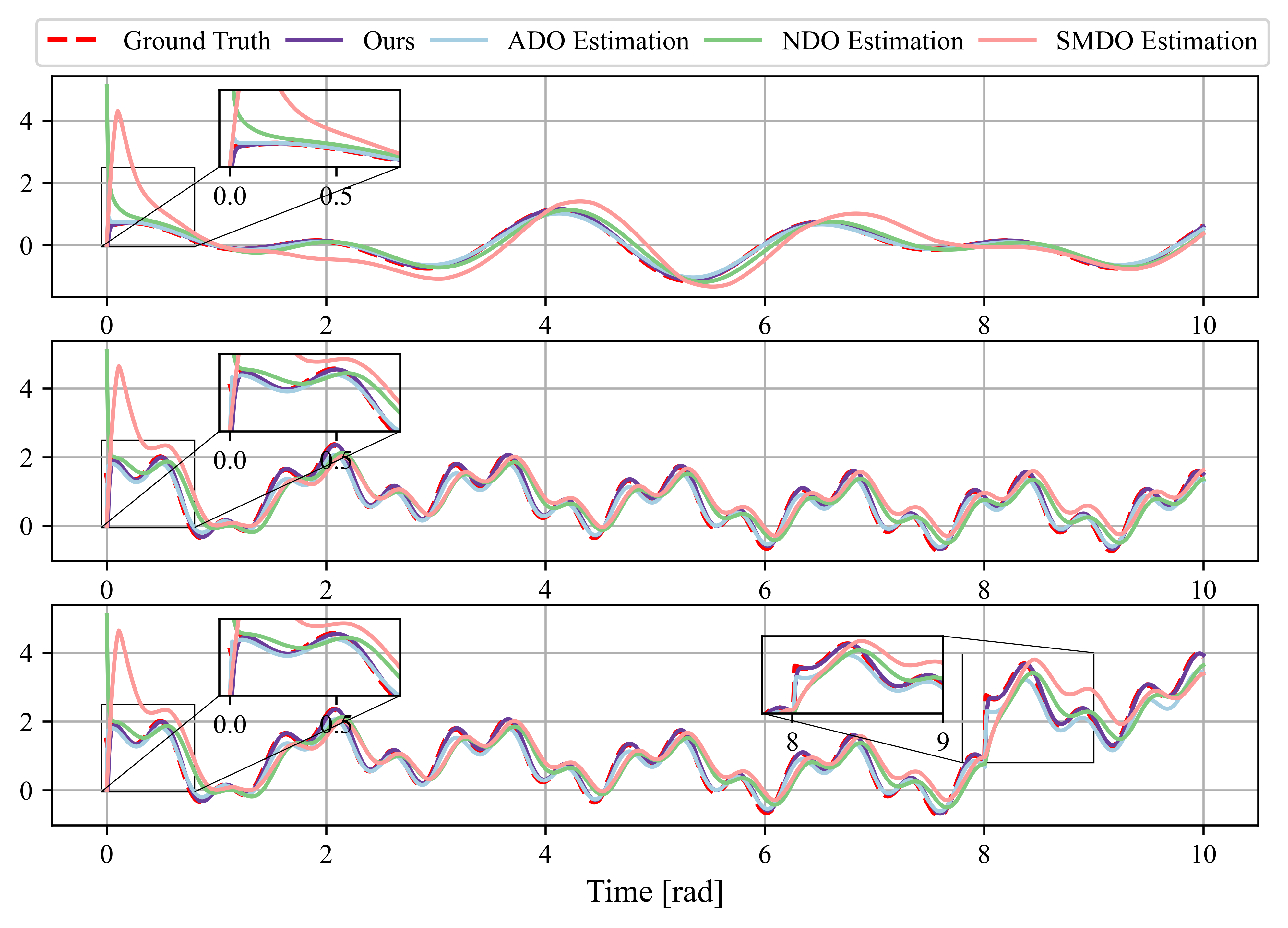}}
	\caption{The estimation results of three disturbance scenarios. From top to bottom, they are $\textit{A}(t)$, $\textit{B}(t,\bm{x})$ and $\textit{C}(t,\bm{x})$, respectively }
	\label{tsr_est_res_C_no_d_com}
\end{figure}

\textit{Observer-only evaluation}: Each observer runs in parallel with the nominal controller but its disturbance estimation is \emph{not} compensated to the control law. That is, the closed-loop control input uses the nominal controler
\begin{equation}
	u = \bm{\pi}(\bm{x}) 
\end{equation} 
The objective is to evaluate the {estimation accuracy} and {convergence characteristics} of each observer independent of any compensation effect on the closed-loop trajectory. The results are given in Fig. \ref{tsr_est_res_C_no_d_com} and Table \ref{rmse_DOs}, which consistently demonstrate that the proposed algorithm outperforms the ADO, NDO, and SMDO across all disturbance scenarios. (1) {Estimation Accuracy}: As shown in Fig. \ref{tsr_est_res_C_no_d_com} and Table \ref{rmse_DOs}, the estimated disturbances produced by our algorithm almost overlap with the ground truth, even for highly nonlinear and state-dependent disturbances. In contrast, ADO and NDO exhibit noticeable steady-state deviations, and SMDO suffers from significant bias and chattering effects. Quantitatively, our approach reduces RMSE by a substantial margin in all cases, demonstrating its superior ability to reconstruct smooth, nonlinear, and discontinuous disturbance patterns with high fidelity. Specifically, compared to ADO, the proposed algorithm reduces the disturbance-estimation RMSE by 43.1\%, 34.7\%, and 51.7\% in the three disturbance scenarios, respectively.  (2) {Convergence Characteristics}: The proposed algorithm also exhibits markedly faster convergence and improved transient response compared with the baseline observers. In the initial phase, the proposed algorithm settles more rapidly without overshoot, while ADO and NDO show delayed convergence and SMDO introduces oscillatory transients. For disturbances containing abrupt transitions ($\textit{C}(t,\bm{x})$), the proposed algorithm captures the changes almost immediately, whereas the baselines display clear phase lag. These results indicate that the proposed algorithm provides not only accurate steady-state estimation but also superior dynamic tracking performance across a wide range of disturbance characteristics. 

\begin{table}[!t]
	\renewcommand{\arraystretch}{1.3}
	\caption{Disturbance Estimation Errors (RMSEs)}
	\centering
	\label{rmse_DOs}
	\begin{tabular}{lcc}
		\toprule
		\textbf{Disturbance Scenario} & \textbf{Observer} & \textbf{Estimation Errors}  \\
		\midrule
		
		\multirow{4}{*}{$A(t)= 0.6(\cos(3t)+\sin(2t))$}
		& ADO & $ 0.065$     \\
		& NDO &$0.239$     \\
		& SMDO &$0.565$     \\
		& Ours &$\textbf{0.034}$     \\
		
		\midrule
		
		\multirow{4}{*}{$\begin{aligned}
				&\textit{B}(\bm{x},t) = 0.5(\cos(12t)+\sin(4t)\\&+\sin(4x_2) +\cos(x_4)+\sin(x_4))
			\end{aligned}$}
		& ADO & $0.144$     \\
		& NDO &$0.382$     \\
		& SMDO &$0.547$     \\
		& Ours &$\textbf{0.087}$     \\
		
		\midrule
		
		\multirow{4}{*}{$C(t,\bm{x})=\textit{B}(\bm{x},t) + 0.8 H(t - t_s)$}
		& ADO & $ 0.205$     \\
		& NDO &$0.414$     \\
		& SMDO &$ 0.598$     \\
		& Ours &$\textbf{0.094}$     \\
		
		\bottomrule
	\end{tabular}
\end{table}

\begin{figure}[!t]
	\centering
	{\includegraphics[width=21pc]{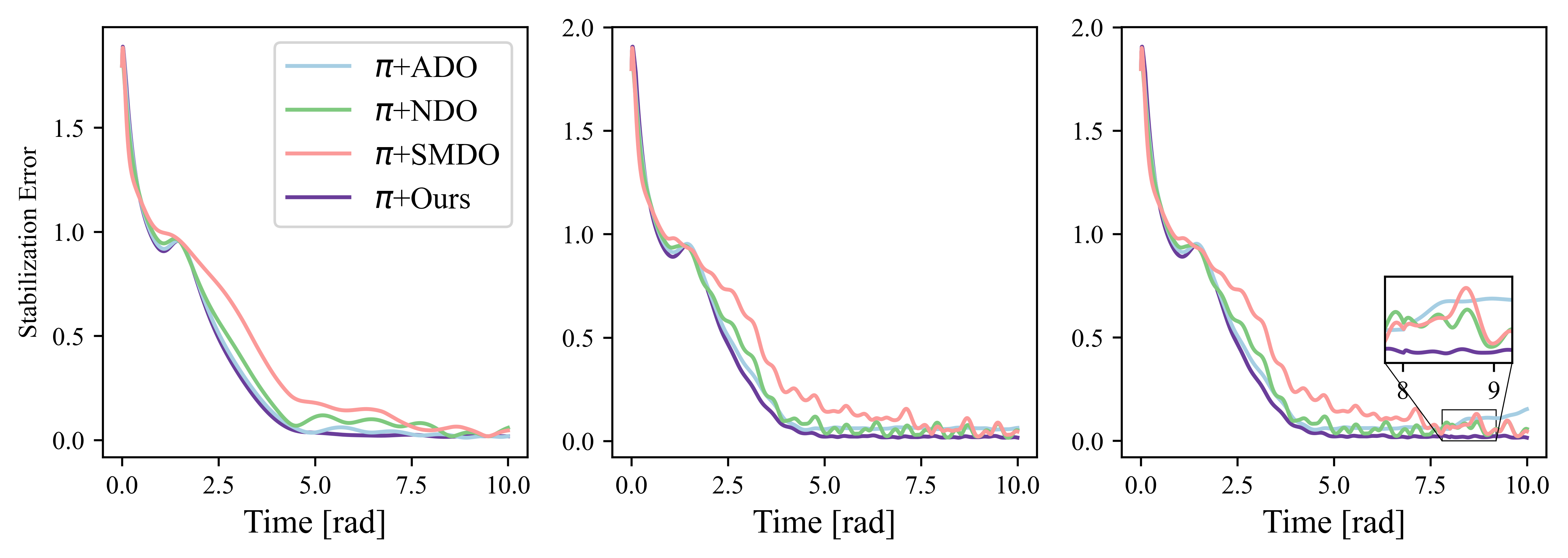}}
	\caption{The stabilization error of three disturbance scenarios. From left to right, they are $\textit{A}(t)$, $\textit{B}(t,\bm{x})$ and $\textit{C}(t,\bm{x})$, respectively }
	\label{tsr_stab_err_C_all}
\end{figure}

\textit{Observer-based compensation}: The estimation of each observer is actively used to compensate in the control law. That is, the closed-loop control input is 
\begin{equation}
	u = \bm{\pi}(\bm{x}) - \hat{\bm{d}}
\end{equation}
which aims to cancel the unknown disturbance. The results of stabilization error is shown in Fig. \ref{tsr_stab_err_C_all}. The stabilization-error trajectories in Fig. \ref{tsr_stab_err_C_all} show that all observers improve the closed-loop regulation performance when their disturbance estimates are fed back for compensation. However, the proposed algorithm achieves the most rapid reduction of the state energy $||\bm{x}||^2$, and reaches the smallest steady-state error across all disturbance scenarios. For the disturbance scenario $\textit{A}(t)$, all observers converge, but our algorithm attains the lowest residual error and the least oscillatory behavior. In the disturbance scenario $\textit{B}(t,\bm{x})$, ADO and NDO exhibit increased transient fluctuations and slower asymptotic decay, whereas our algorithm maintains a monotonic convergence trend with minimal variation. The advantage becomes even more pronounced for the step-affected disturbance $\textit{C}(t,\bm{x})$. After the abrupt disturbance change, SMDO generates large oscillations and NDO/ADO display prolonged recovery phases with noticeable stabilization error. In contrast, the proposed algorithm rapidly suppresses the error and stabilizes at a significantly lower level, indicating its stronger robustness against discontinuities and nonlinear disturbance components. 

Overall, the numerical simulation results indicate that: (1) The proposed algorithm provides not only accurate steady-state estimation but also superior dynamic tracking performance across a wide range of disturbance characteristics; (2) Incorporating the proposed algorithm into the control loop yields faster stabilization, improved disturbance rejection, and superior steady-state regulation compared with the baseline observers.

\section{Numerical Evaluation-B}\label{section_simulation_B}
In this section, to further validate the superiority of the proposed algorithm, we conduct comparative numerical simulations against two state-of-the-art learning-based estimators, namely NeuroBEM \cite{Bauersfeld2021} and NeuroMHE \cite{Wang2024e}. NeuroBEM employs a neural-network-based basis expansion to approximate unknown system dynamics, while NeuroMHE integrates neural networks with moving horizon estimation for joint state and disturbance estimation. The NeuroBEM dataset\footnote{\url{https://rpg.ifi.uzh.ch/NeuroBEM.html}} is adopted in this section because it provides a standardized benchmark with measured quadrotor states, enabling a reliable and quantitative evaluation of external forces and torques estimation accuracy.

\subsection{Dynamics of Quadrotor}
It is assumed that the quadrotor is a 6 degree-of-freedom (DoF) rigid body with mass $m$ and moment of inertia $\bm{J}\in \mathbb{R}^{3\times3}$. Then the dynamics of quadrotor could be written as 
\begin{equation}
	\begin{aligned}
		&\dot{\boldsymbol{p}}_w=\boldsymbol{v}_w, \ 
		\dot{\boldsymbol{v}}_w=m^{-1}\left(-mg\bm{z}+\boldsymbol{R}f_u\bm{z}+\bm{f}_{\text{e}}\right)\\
		&\dot{\bm{R}}=\bm R \bm \omega_b^{\times }, \ 
		\dot{\boldsymbol{\omega}_b}=\boldsymbol{J}^{-1}\left(-\bm \omega_b^{\times}\bm J \bm \omega_b+\boldsymbol{\tau}_m+ \bm{\tau}_{\text{e}}\right) 
	\end{aligned}\label{quadrotordynamics}
\end{equation}
where $\bm{f}_{\text{e}}$ and $\bm{\tau}_{\text{e}}$ are external residual forces and torques, respectively. And $\bm{p}_w=\left[x,y,z\right]^\mathrm{T}$ and $\bm{v}_w=\left[v_x,v_y,v_z\right]^T$ are the position and velocity of the quadrotor's center-of-mass in the world frame $\mathcal{I}$, $\boldsymbol{\omega}_b=\left[{\omega}_x,{\omega}_y,{\omega}_z\right]^\mathrm{T}$ is the angular velocity in  the quadrotor body frame $\mathcal{B}$. The $\boldsymbol{R}\in SO^3$ is the rotation matrix from $\mathcal{B}$ and $\mathcal{I}$. The gravitational acceleration is $g$ and $\bm{z}=\left[0,0,1\right]^\mathrm{T}$. The $f_u$ and $\boldsymbol{\tau}_m=\left[{\tau}_{mx},{\tau}_{my},{\tau}_{mz}\right]^\mathrm{T}$ denote the total thrust and torque produced by the quadrotor’s four motors. We define $\bm{x}=\left[\bm{p}_w^\mathrm{T},\bm{v}_w^\mathrm{T},vec(\bm R),\bm{\omega}^\mathrm{T}\right]^\mathrm{T}$ as the quadrotor state, where $vec(\cdot)$ represents the vectorization of matrix and $\bm{u} = [{f}_u, {\tau}_{mx},{\tau}_{my},{\tau}_{mz}]^\mathrm{T}$ as the control input, respectively.

\subsection{Implementation Details}
\textit{Simulation Setup}: As the external forces and torques differ significantly in magnitude, two separate instances of the proposed algorithm are deployed: one dedicated to estimating the residual external forces and the other to estimating the residual external torques\footnote{This separation avoids scale-induced bias during learning and allows each instance to be tuned according to the corresponding signal characteristics.}. Both instances share the same network architecture and lifting functions design. The instance used to capture the external forces takes the linear velocity of quadrotor as its input, e.g. $\bm{\zeta}=[\bm{f}_e\ \bm{v}]$. The instance used to capture the external torques takes the angular velocity of quadrotor as its input, e.g. $\bm{\zeta}=[\bm{\tau}_e\ \bm{w}_b]$. The architecture of the lifting functions $\bm{\psi}$ is identical to those described in Section \ref{section_simulation_A}, but the dimension of $\bm{\psi}$ is set to 32, i.e. $N=32$. The learning rate is 0.001.

\textit{Evaluation Metrics}: We evaluate the performance of NeuroBEM, NeuroMHE and the proposed algorithm in terms of external forces estimation accuracy $\text{RMSE}(\hat{\bm{f}}_e,\bm{f}_e)$ and torques estimation accuracy $\text{RMSE}(\hat{\bm{\tau}}_e,\bm{\tau}_e)$.

\begin{table*}[htbp]
	\centering
	\caption{Comparison results of estimation errors (RMSEs) on 12 agile flight trajectories. The best results (lowest errors) are highlighted in bold.}
	\label{bem_trj_comparison}

	\resizebox{\textwidth}{!}{%
		\begin{minipage}[t]{0.54\linewidth}
			\centering
			\begin{tabular}{cccccccc}
				\toprule
				\textbf{Traj} & \textbf{Method} & $\boldsymbol{f_x}$ & $\boldsymbol{f_y}$ & $\boldsymbol{f_z}$ & $\boldsymbol{\tau_x}$ & $\boldsymbol{\tau_y}$ & $\boldsymbol{\tau_z}$ \\
				\midrule
				
				\multirow{3}{*}{\shortstack{3D\\Circle\_1}} 
				& NeuroBEM & 0.196 & 0.211 & 0.215 & 0.005 & 0.006 & 0.003 \\
				& NeuroMHE & 0.258 & 0.269 & 0.108 & 0.003 & 0.002 & 0.003 \\
				& DEKC     & \textbf{0.079} & \textbf{0.051} & \textbf{0.075} & \textbf{0.002} & \textbf{0.001} & \textbf{0.002} \\
				\midrule
				
				\multirow{3}{*}{\shortstack{Linear\\Oscill.}} 
				& NeuroBEM & 0.164 & 0.185 & 0.456 & 0.013 & 0.011 & 0.006 \\
				& NeuroMHE & \textbf{0.119} & 0.105 & 0.186 & 0.011 & 0.007 & 0.005 \\
				& DEKC     & 0.130 & \textbf{0.036} & \textbf{0.109} & \textbf{0.008} & \textbf{0.005} & \textbf{0.004} \\
				\midrule
				
				\multirow{3}{*}{\shortstack{Race\\Track\_1}} 
				& NeuroBEM & 0.169 & 0.158 & 0.463 & 0.009 & 0.009 & \textbf{0.004} \\
				& NeuroMHE & 0.141 & 0.092 & 0.115 & 0.007 & 0.004 & \textbf{0.004} \\
				& DEKC     & \textbf{0.100} & \textbf{0.035} & \textbf{0.094} & \textbf{0.006} & \textbf{0.003} & 0.006 \\
				\midrule
				
				\multirow{3}{*}{\shortstack{Race\\Track\_2}} 
				& NeuroBEM & 0.262 & 0.248 & 0.552 & 0.014 & 0.012 & 0.007 \\
				& NeuroMHE & 0.245 & 0.175 & 0.208 & 0.012 & 0.008 & 0.018 \\
				& DEKC     & \textbf{0.152} & \textbf{0.053} & \textbf{0.154} & \textbf{0.009} & \textbf{0.005} & \textbf{0.006} \\
				\midrule
				
				\multirow{3}{*}{\shortstack{3D\\Circle\_2}} 
				& NeuroBEM & \textbf{0.110} & 0.129 & 0.470 & 0.006 & 0.009 & 0.004 \\
				& NeuroMHE & 0.140 & 0.135 & \textbf{0.075} & 0.003 & 0.002 & 0.004 \\
				& DEKC     & 0.148 & \textbf{0.044} & 0.119 & \textbf{0.002} & \textbf{0.001} & \textbf{0.002} \\
				\midrule
				
				\multirow{3}{*}{\shortstack{Figure\\-8\_2}} 
				& NeuroBEM & 0.051 & 0.036 & 0.339 & 0.002 & 0.002 & 0.002 \\
				& NeuroMHE & \textbf{0.020} & 0.058 & 0.029 & 0.002 & \textbf{0.001} & 0.002 \\
				& DEKC     & \textbf{0.020} & \textbf{0.017} & \textbf{0.022} & \textbf{0.001} & \textbf{0.001} & \textbf{0.001} \\
				\bottomrule
			\end{tabular}
		\end{minipage}
		\hfill 
		\begin{minipage}[t]{0.54\linewidth}
			\centering
			\begin{tabular}{cccccccc}
				\toprule
				\textbf{Traj} & \textbf{Method} & $\boldsymbol{f_x}$ & $\boldsymbol{f_y}$ & $\boldsymbol{f_z}$ & $\boldsymbol{\tau_x}$ & $\boldsymbol{\tau_y}$ & $\boldsymbol{\tau_z}$ \\
				\midrule
				
				\multirow{3}{*}{\shortstack{Melon\_1}} 
				& NeuroBEM & 0.099 & 0.108 & 0.397 & 0.004 & 0.005 & 0.003 \\
				& NeuroMHE & \textbf{0.053} & 0.059 & 0.060 & 0.003 & \textbf{0.001} & \textbf{0.002} \\
				& DEKC     & 0.055 & \textbf{0.027} & \textbf{0.041} & \textbf{0.002} & \textbf{0.001} & \textbf{0.002} \\
				\midrule
				
				\multirow{3}{*}{\shortstack{Figure\\-8\_3}} 
				& NeuroBEM & 0.145 & 0.168 & 0.584 & \textbf{0.010} & 0.012 & 0.006 \\
				& NeuroMHE & \textbf{0.118} & 0.133 & 0.151 & \textbf{0.010} & 0.006 & 0.005 \\
				& DEKC     & 0.132 & \textbf{0.070} & \textbf{0.143} & \textbf{0.010} & \textbf{0.004} & \textbf{0.004} \\
				\midrule
				
				\multirow{3}{*}{\shortstack{Figure\\-8\_4}} 
				& NeuroBEM & 0.400 & 0.313 & 1.084 & 0.020 & 0.018 & 0.009 \\
				& NeuroMHE & \textbf{0.169} & 0.174 & 0.237 & \textbf{0.010} & 0.010 & \textbf{0.007} \\
				& DEKC     & 0.199 & \textbf{0.095} & \textbf{0.188} & 0.012 & \textbf{0.006} & \textbf{0.007} \\
				\midrule
				
				\multirow{3}{*}{\shortstack{Melon\_2}} 
				& NeuroBEM & 0.244 & 0.198 & 0.921 & 0.009 & 0.006 & \textbf{0.003} \\
				& NeuroMHE & 0.254 & 0.213 & \textbf{0.094} & 0.005 & 0.003 & 0.004 \\
				& DEKC     & \textbf{0.186} & \textbf{0.059} & 0.278 & \textbf{0.003} & \textbf{0.002} & \textbf{0.003} \\
				\midrule
				
				\multirow{3}{*}{\shortstack{Random\\Points}} 
				& NeuroBEM & 0.161 & 0.183 & 0.471 & 0.008 & 0.008 & 0.005 \\
				& NeuroMHE & 0.115 & 0.114 & 0.204 & 0.010 & 0.006 & 0.005 \\
				& DEKC     & \textbf{0.068} & \textbf{0.041} & \textbf{0.084} & \textbf{0.007} & \textbf{0.003} & \textbf{0.003} \\
				\midrule
				
				\multirow{3}{*}{\shortstack{Ellipse}} 
				& NeuroBEM & 0.204 & 0.315 & 1.039 & 0.012 & 0.008 & 0.005 \\
				& NeuroMHE & \textbf{0.176} & 0.165 & \textbf{0.089} & 0.005 & 0.003 & 0.006 \\
				& DEKC     & 0.212 & \textbf{0.078} & 0.256 & \textbf{0.003} & \textbf{0.002} & \textbf{0.003} \\
				\bottomrule
			\end{tabular}
		\end{minipage}
	}
\end{table*}

\subsection{Evaluation Results and Theoretical Discussion}
We use the same 12 agile flight trajectories in \cite{Wang2024e} to evaluate the performance of the proposed algorithm against NeuroBEM and NeuroMHE. The detailed results are provided in Table \ref{bem_trj_comparison}. The proposed algorithm demonstrates statistically significant improvements in external forces/torques estimation accuracy compared to the baseline methods (NeuroBEM and NeuroMHE) across a diverse range of flight trajectories. As evidenced by the quantitative results, the proposed algorithm consistently yields the lowest residuals in both force and torque domains, particularly in scenarios characterized by high aerodynamic nonlinearity and aggressive maneuvering.

{Force Estimation Enhancement}: In terms of external forces estimation, the proposed algorithm exhibits substantial robustness against unmodeled dynamics. This advantage is most pronounced in the $z$-axis during aggressive flights. For instance, in the Random Points" trajectory, DEKC reduces the residual force estimation error in $z$-axis to {0.084 N}, representing a performance improvement of approximately {82.2\%} over NeuroBEM (0.471 N) and {58.8\%} over NeuroMHE (0.204 N). Similarly, in the high-speed Figure-8\_4"\footnote{This trajectory covers a velocity range of  0.01 m/s to 17.72 m/s} trajectory, DEKC successfully suppresses aerodynamic modeling errors, lowering the residual force estimation error in $z$-axis by {82.7\%} compared to NeuroBEM (1.084 N). The improvements are also evident in lateral dynamics. In the ``Race Track\_1" scenario, DEKC reduces the residual force estimation error in $y$-axis by {77.8\%} relative to NeuroBEM and {62.0\%} relative to NeuroMHE, confirming its superior capability in compensating for complex drag and side-slip effects.

{Torque Estimation Fidelity}: In terms of external torques estimation, compared with the other two methods, the proposed DEKC demonstrates superior performance on the majority of trajectories and achieves comparable performance on the remaining ones. In the Ellipse" trajectory, the residual torque estimation error in $x$-axis is minimized to {0.003 Nm}, marking a {75.0\%} improvement over NeuroBEM. Consistent gains are observed in the Melon\_2" trajectory, where DEKC outperforms NeuroMHE by {33.3\%} in residual torque estimation in $y$-axis (0.002 Nm vs 0.003 Nm). Furthermore, in maneuvers such as ``3D Circle\_1," DEKC maintains a clear advantage, reducing residual torque estimation error in $x$-axis by {60.0\%} compared to NeuroBEM and {33.3\%} compared to NeuroMHE.

\begin{figure*}[htbp]
	\centering
	\subfigure[Random Points case.]{\includegraphics[width=21pc]{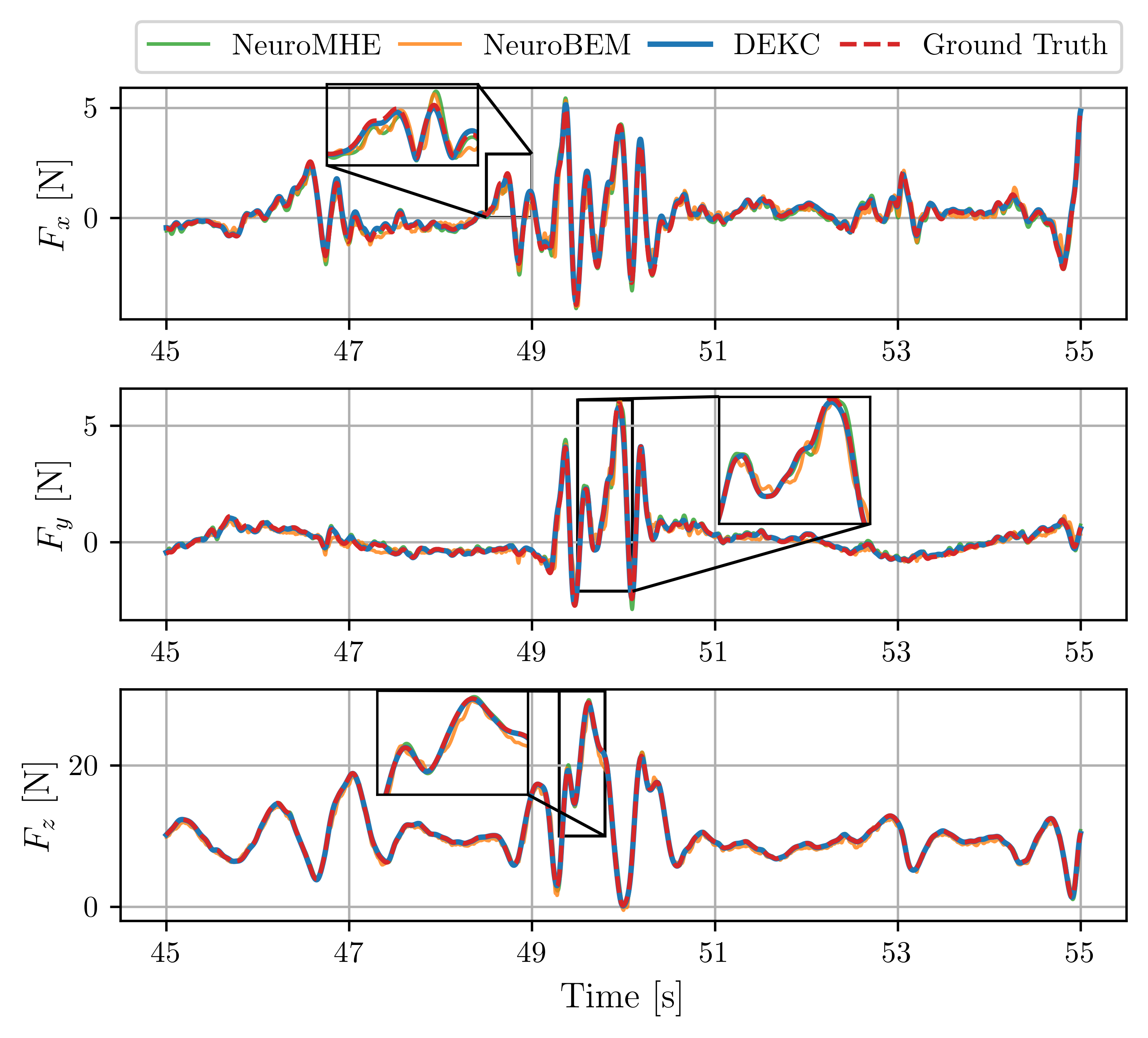}}
	\subfigure[Race Track\_2 case.]{\includegraphics[width=21pc]{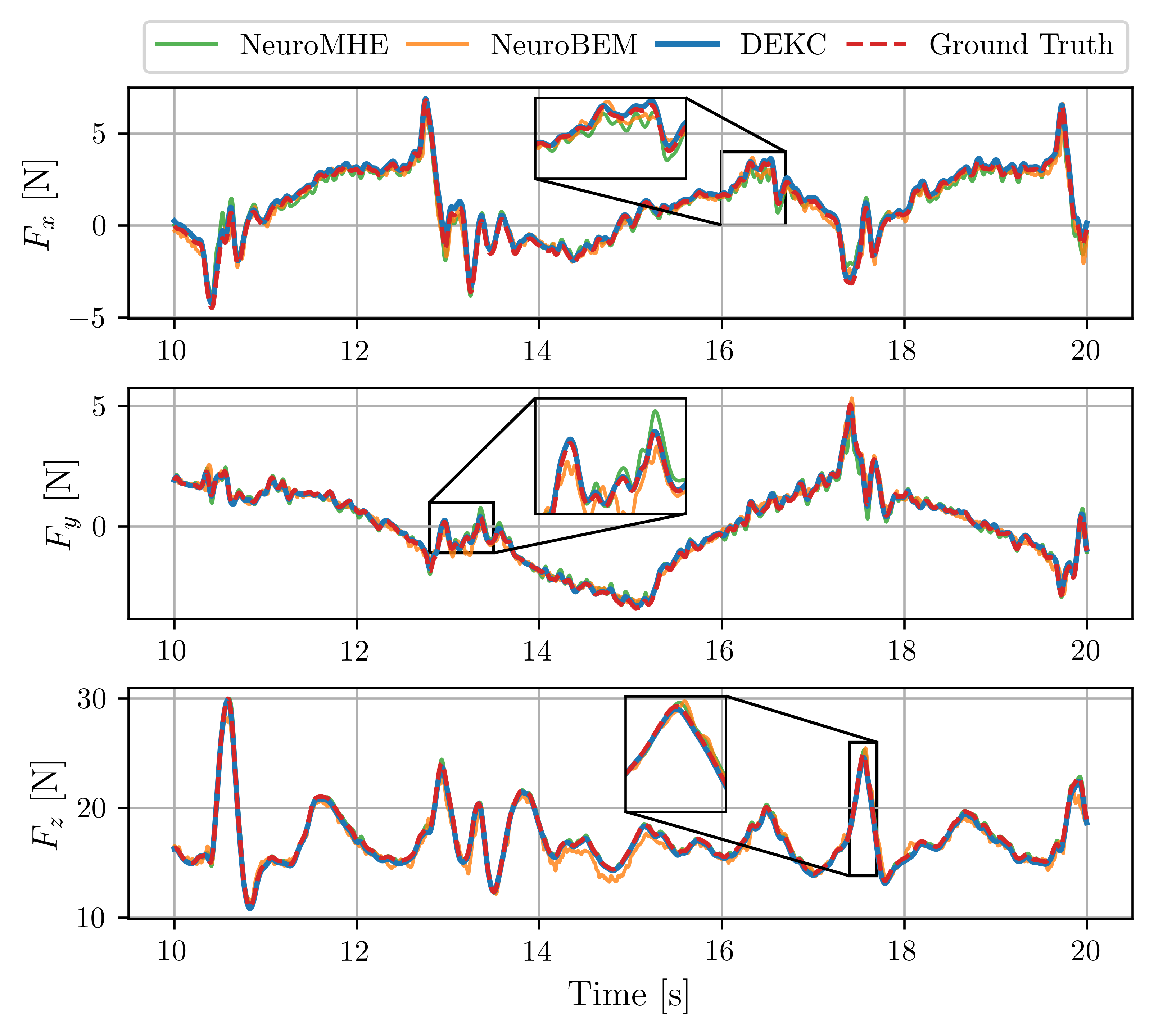}}
	\caption{Force Estimation performance of DEKC, NeuroBEM and NeuroMHE on two test datasets. The $F_x$, $F_y$ and $F_z$ denote the external forces acting on the quadrotor, expressed in the body frame. The proposed DEKC (blue solid line) demonstrates superior estimation performance compared to the baselines.}
	\label{bem_comparision}
\end{figure*}

In addition, the comparative evaluation of the force estimation performance is visually depicted in Fig. \ref{bem_comparision}, which contrasts the proposed DEKC method against the NeuroBEM and NeuroMHE baselines across the ``Random Points" and ``Race Track\_2" datasets. The visual evidence corroborates the quantitative improvements presented in Table \ref{bem_trj_comparison}, highlighting the superior performance of DEKC in handling both highly aggressive transients and complex aerodynamic disturbances. Fig. \ref{bem_comparision}(a) illustrates the estimation results on the "Random Points" dataset. As evidenced by the zoomed-in views of the $F_{x}$, $F_y$ and $F_z$ components, the baseline methods exhibit noticeable performance degradation. Specifically, NeuroBEM tends to suffer from aggressive maneuvering, indicating a potential sensitivity to unmodeled high-frequency aerodynamics. Conversely, NeuroMHE exhibits a slight phase lag and attenuation in peak magnitude, likely due to the trade-off between the horizon length and transient responsiveness. In contrast, the estimation of DEKC is closest to the Ground Truth, demonstrating a faster convergence rate and an enhanced capability to capture rapid changes in external forces without inducing significant oscillations. Furthermore, the superior performance against aerodynamic modeling errors is pronounced in the high-speed "Race Track\_2" scenario shown in Fig. \ref{bem_comparision}(b). In this regime, unmodeled drag effects become dominant, particularly in the $z$-axis. The enlarged view of the $F_z$ component reveals a distinct steady-state bias in the NeuroBEM estimation, suggesting that the pre-trained model may struggle to generalize to velocities outside the training distribution. However, DEKC successfully compensates for these complex aerodynamic discrepancies, effectively eliminating the residual error. Additionally, although NeuroMHE captures the general aerodynamic trend better than NeuroBEM, it fails to achieve the precise convergence seen in DEKC.

The above results indicate that DEKC demonstrates a statistically significant advancement over both NeuroBEM and NeuroMHE, consistently delivering superior forces/torques estimation accuracy, particularly within high-dynamic and aggressive flight scenarios. The core distinction among these three methods lies in the fundamental residual dynamics modeling paradigm. NeuroBEM and NeuroMHE treat the residual dynamics as a functional mapping (regression) or an optimization parameter, whereas DEKC treats it as a dynamical system with its own learnable evolution. The detailed analysis of three methods is presented as follows.

The NeuroBEM employs a temporal-convolutional network (TCN) to map a sliding window of state-control history to the residual dynamics, i.e., $\bm{d}_k \approx \mathcal{NN}(\bm{x}_{k-h:k}, \bm{u}_{k-h:k})$\footnote{The $\bm{x}_{k-h:k}$ and $\bm{u}_{k-h:k}$ denote the state and input sequences over the horizon from $k-h$ to $k$, respectively, where $\bm{x}_{k-h:k} := [\bm{x}_{k-h}, \ldots, \bm{x}_k]$ and $\bm{u}_{k-h:k} := [\bm{u}_{k-h}, \ldots, \bm{u}_k]$.}. Although this architecture can captures short-term temporal dependencie, such as rotor wake interactions and blade flapping, it fundamentally remains a black-box regression approach. The neural network learns an implicit representation of the residual dynamics. Consequently, when the quadrotor executes unstructured or aggressive maneuvers (e.g., the ``Random Points" trajectory) that deviate from the distribution during training, the NeuroBEM lacks a mechanism to enforce physical consistency. It essentially attempts to "match patterns" in the input history, leading to significant prediction degradation (e.g., residual force estimation error in $z$-axis of ``Figure-8\_4" is 1.084 N) when the input sequence lies out-of-distribution.

The NeuroMHE framework embeds neural networks within a Moving Horizon Estimation (MHE) scheme, formulating the estimation of residual dynamics as a constrained optimization problem. While exhibiting robustness against measurement noise, the approach inherently entails a trade-off between estimation accuracy and the chosen horizon length. The "backward-looking" nature of the sliding window essentially acts as a smoothing filter, inducing phase lag during rapid directional changes. Furthermore, NeuroMHE typically models disturbances as bounded parameters to be fitted within the window rather than dynamical states with predictable evolution, limiting its ability to anticipate drastic aerodynamic fluctuations.

The superior performance of the DEKC algorithm can be attributed to its fundamental departure from the regression-based paradigm of NeuroBEM and the optimization-based estimation of NeuroMHE. The DEKC algorithm fundamentally shifts the problem from fitting or optimization to learning the evolution of residual dynamics. DEKC employs the DDKO theory to explicitly model residual dynamcis as a dynamic process governed by a linear Koopman operator in a lifted space. By learning the evolution operator that propagates the residual dynamics forward in time, DEKC effectively captures the underlying "physics" of the residual dynamics, such as the decay rates and oscillatory modes of aerodynamic drag. This transforms the estimation task from a reactive compensation (fitting the current error) to a predictive tracking problem. Consequently, DEKC achieves significantly lower latency and higher accuracy in both force and torque estimation, as it accounts for the temporal coherence of residual dynamics that baseline methods fail to resolve.

\subsection{Computational Efficiency}
To evaluate the real-time feasibility of the proposed DEKC framework, we evaluated the average computation time per control cycle using the NeuroBEM dataset. The tests were performed on a Linux workstation equipped with Intel Core i9-12900H and 16G RAM. The average computation time of DEKC is 1.35 ms (the standard deviation is 0.58 ms). This implies a potential update rate of approximately 500 Hz. 

To further assess the algorithm's deployability on resource-constrained onboard platform, we benchmarked the proposed DEKC framework on an embedded computer (equipped with an RK3588 chip). The average computation time was recorded at 5 ms, corresponding to an update frequency of 200 Hz. Given that standard quadrotor flight control loops typically operate within the 100–500 Hz range, this latency confirms that DEKC is computationally efficient enough for real-time onboard execution. In contrast, optimization-based methods such as NeuroMHE often necessitate high-performance desktop-grade CPUs to maintain comparable update rates.

\section{Real-World Validation}\label{section_experiment}
In this section, we apply the proposed DEKC-MPC algorithm to the flight control of quadrotor under various uncertainties. To comprehensively evaluate the robustness and adaptability of the framework, we designed three distinct experimental scenarios: (1) Trajectory Tracking, where the quadrotor follows a complex 3D trajectory to verify basic tracking performance; (2) Payload Transportation, where the quadrotor tracks the 3D trajectory while carrying a suspended payload, introducing significant internal model mismatch; and (3) Hovering under Compound Disturbances, where the quadrotor is subjected to both the suspended payload and external wind gusts simultaneously, testing the limits of the disturbance rejection capability.

\subsection{Experiments Setup}
\begin{figure}[!t]
	\centering
	{\includegraphics[width=21pc]{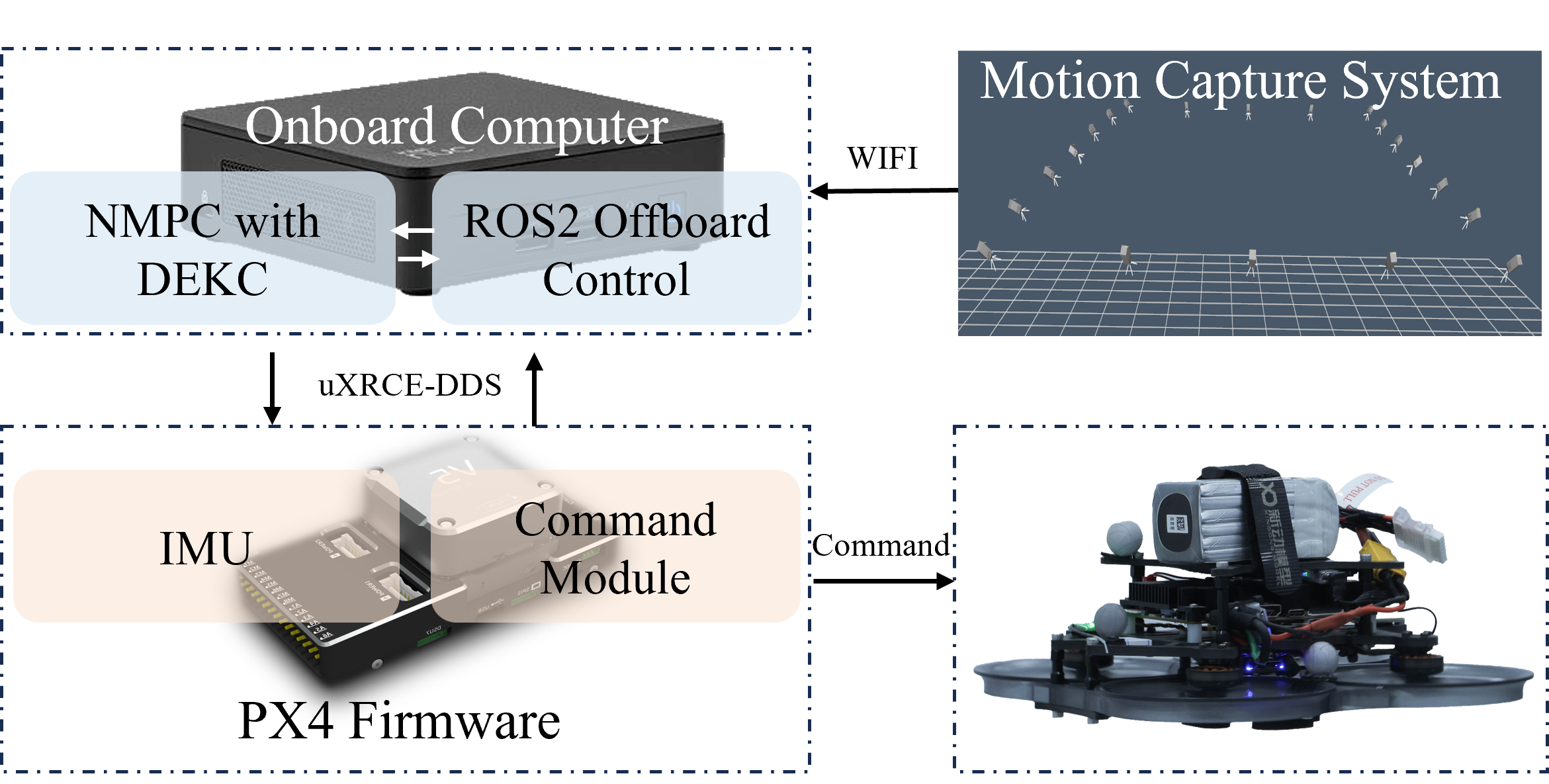}}
	\caption{Hardware and software architecture of the experimental platform. The system adopts a dual-loop control structure. The onboard computer (top left) runs the high-level ROS2 nodes for DEKC estimation and NMPC optimization, receiving ground truth of position from the motion capture system via WiFi. Control commands are sent via uXRCE-DDS to the flight controller running PX4 Firmware (bottom left) for low-level execution. The custom-built quadrotor is shown on the bottom right.}
	\label{experiment_setup}
\end{figure}

The setup for real-world validation is shown in Fig. \ref{experiment_setup}. The custom-built quadrotor weighs 0.65 kg, with a motor wheelbase of 155 mm. The low-level control and attitude estimation is managed by a flight controller, running the PX4 autopiot firmware (version 1.15.4). The real-time position and attitude of quadrotor are obtained from a Nokov motion capture system operating at 100 hz. CasADi \cite{Andersson2019} and Acados \cite{Verschueren2022} are employed for solving the optimization problem (\ref{mpc_dekc}). The predictive horizon is se to 50 and the the optimization problem (\ref{mpc_dekc}) is solved at 50Hz on the onboard companion computer. The architecture of $\bm{\psi}_{\theta}$ is identical to those described in Section \ref{section_simulation_A}, but the dimension of $\bm{\psi}_{\theta}$ is set to 32, i.e. $N=32$. The learning rate is 0.01.

\subsection{Comparative Baseline Methods}
To rigorously evaluate the performance of the proposed DEKC-MPC framework, we conduct comparative experiments against two representative control schemes: a standard model-based approach (Nominal MPC) and a traditional adaptive approach (LR-MPC). 

The Nominal MPC represents the standard NMPC formulation that relies solely on the a priori nominal model of the quadrotor. It assumes the external disturbance is zero (i.e., $\hat{d}_k = 0$) throughout the prediction horizon. This scheme serves as a performance baseline to quantify the degradation of control accuracy caused by model mismatch and external disturbances when no compensation mechanism is applied. 

Linear Regression-Augmented MPC (LR-MPC) is implemented by augmenting the Nominal-MPC with an online Linear Regression (LR) disturbance observer. This method assumes that the external disturbance can be linearly parameterized by a set of hand-crafted basis functions 
\begin{equation}
	\hat{d}_k = W_k^\top \chi(x_k)
\end{equation}
where $\chi(x)$ is the function of velocity. The weight matrix $W_k$ is updated online using a sliding-window least squares algorithm with a window length of 50 to minimize the fitting error. This baseline represents the class of physics-inspired adaptive controllers. By comparing DEKC-MPC with LR-MPC, we aim to verify whether the learnable deep embedding of the Koopman operator provides superior expressiveness and generalization capability over fixed, manually designed basis functions, especially under complex, nonlinear disturbance conditions.

\begin{table}[!t]
	\centering 
	\renewcommand{\arraystretch}{1.4}
	\caption{RMSE Results of Tracking Performance on Five Flight Trajectories} \label{table_experiment_scene_a}
	\begin{tabular}{lcccc} 
		\toprule
		\textbf{Trajectory}  & \textbf{Method} & \textbf{$E_x$} [m]  & \textbf{$E_y$} [m] & \textbf{$E_z$} [m] \\ 
		\midrule
		
		\multirow{3}{*}{\shortstack{3D Loop \\ \includegraphics[width=1.5cm]{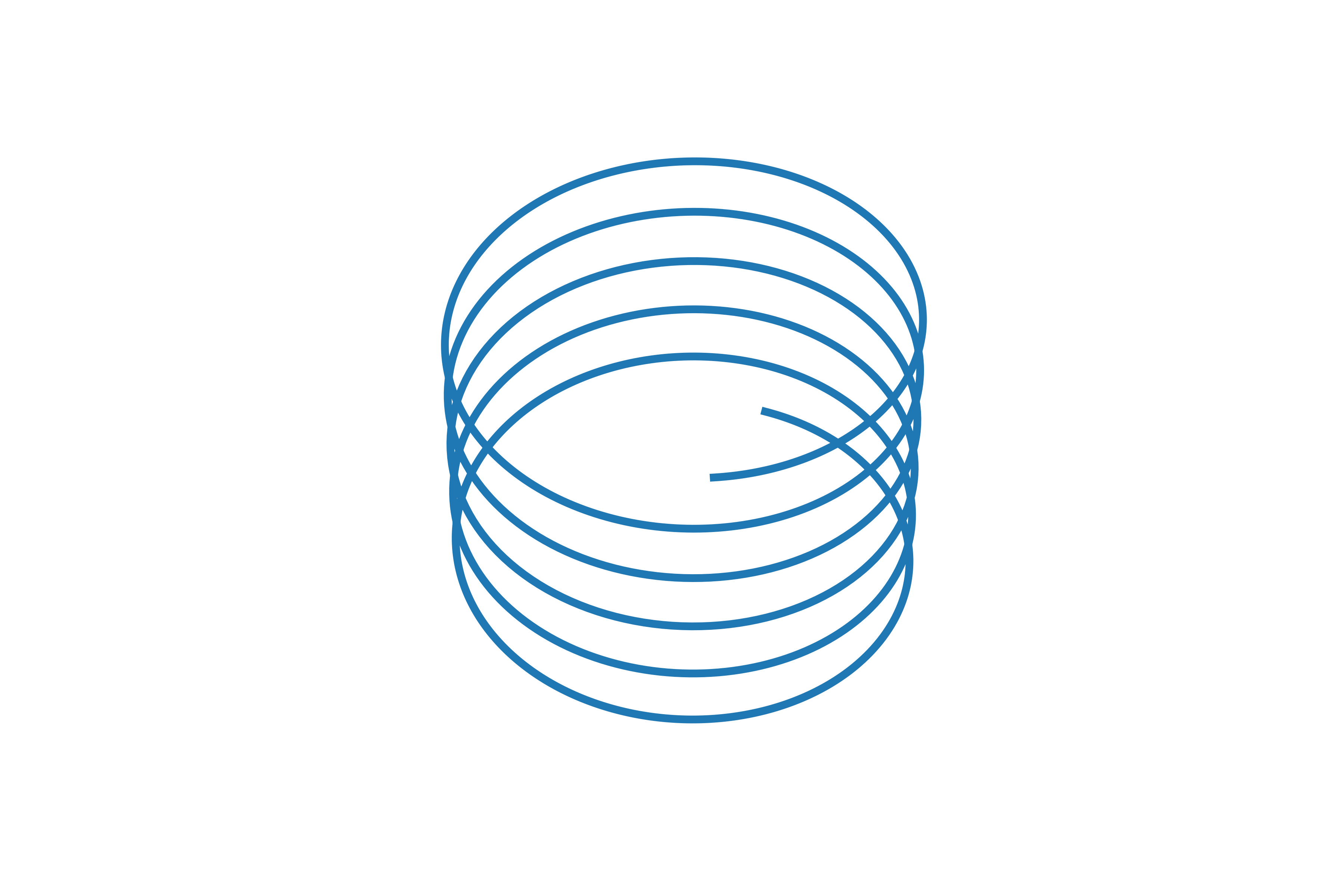}}}  
		& Nominal-MPC &  0.0789 &0.0704 & 0.1188      \\
		& LR-MPC & 0.0646 & 0.0645  & 0.0843   \\
		& DEKC-MPC &\textbf{ 0.0576} & \textbf{0.0590} & \textbf{0.0522}    \\
		\midrule
		
		\multirow{3}{*}{\shortstack{Conical Spiral \\ \includegraphics[width=1.5cm]{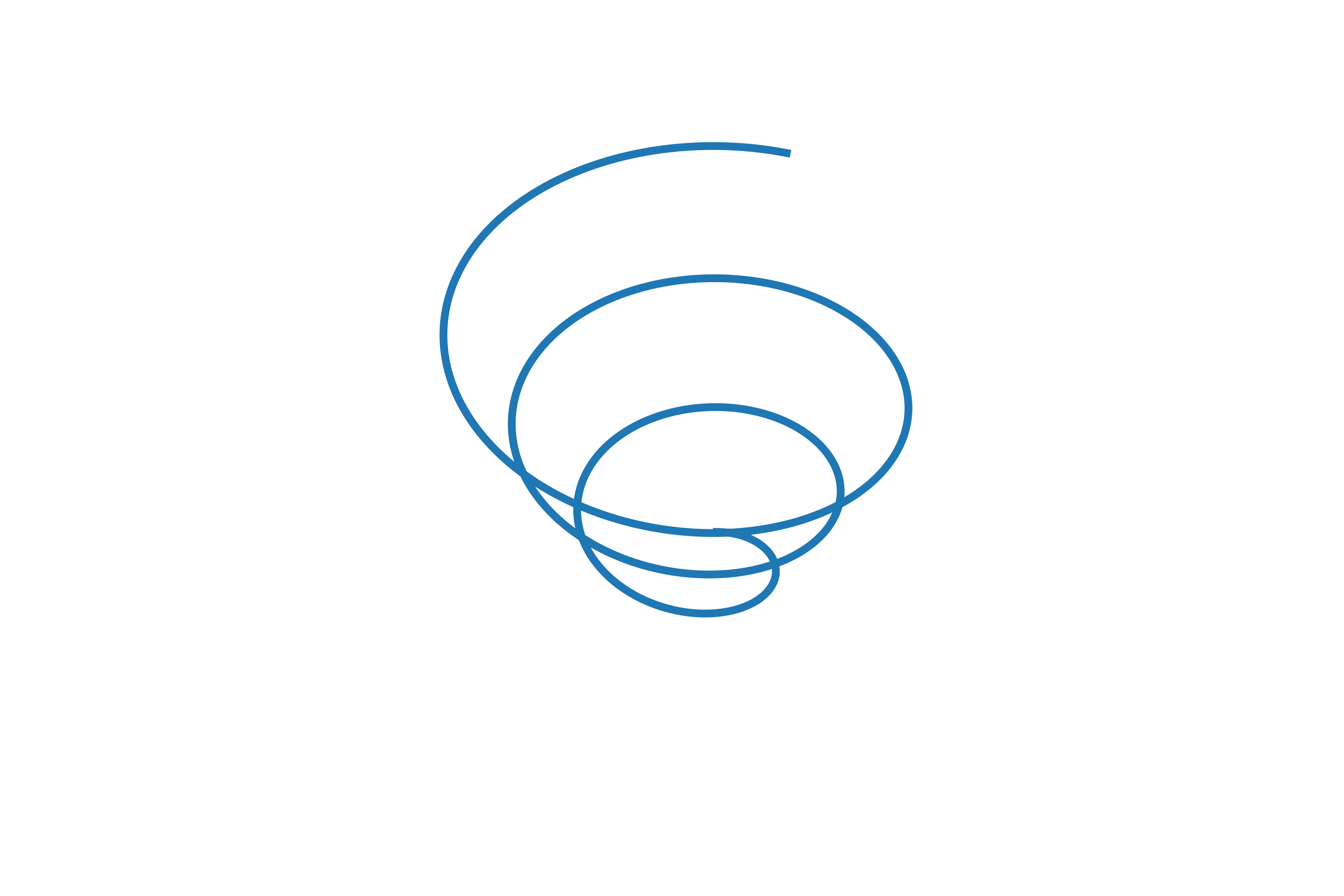}}} 
		& Nominal-MPC & 0.0761 & 0.0655 & 0.0719 \\
		& LR-MPC      & 0.0654 & 0.0566 & 0.0567 \\
		& DEKC-MPC    & \textbf{0.0548} & \textbf{0.0485} & \textbf{0.0357} \\ 
		\midrule	
		
		\multirow{3}{*}{\shortstack{Lemniscate \\ \includegraphics[width=1.5cm]{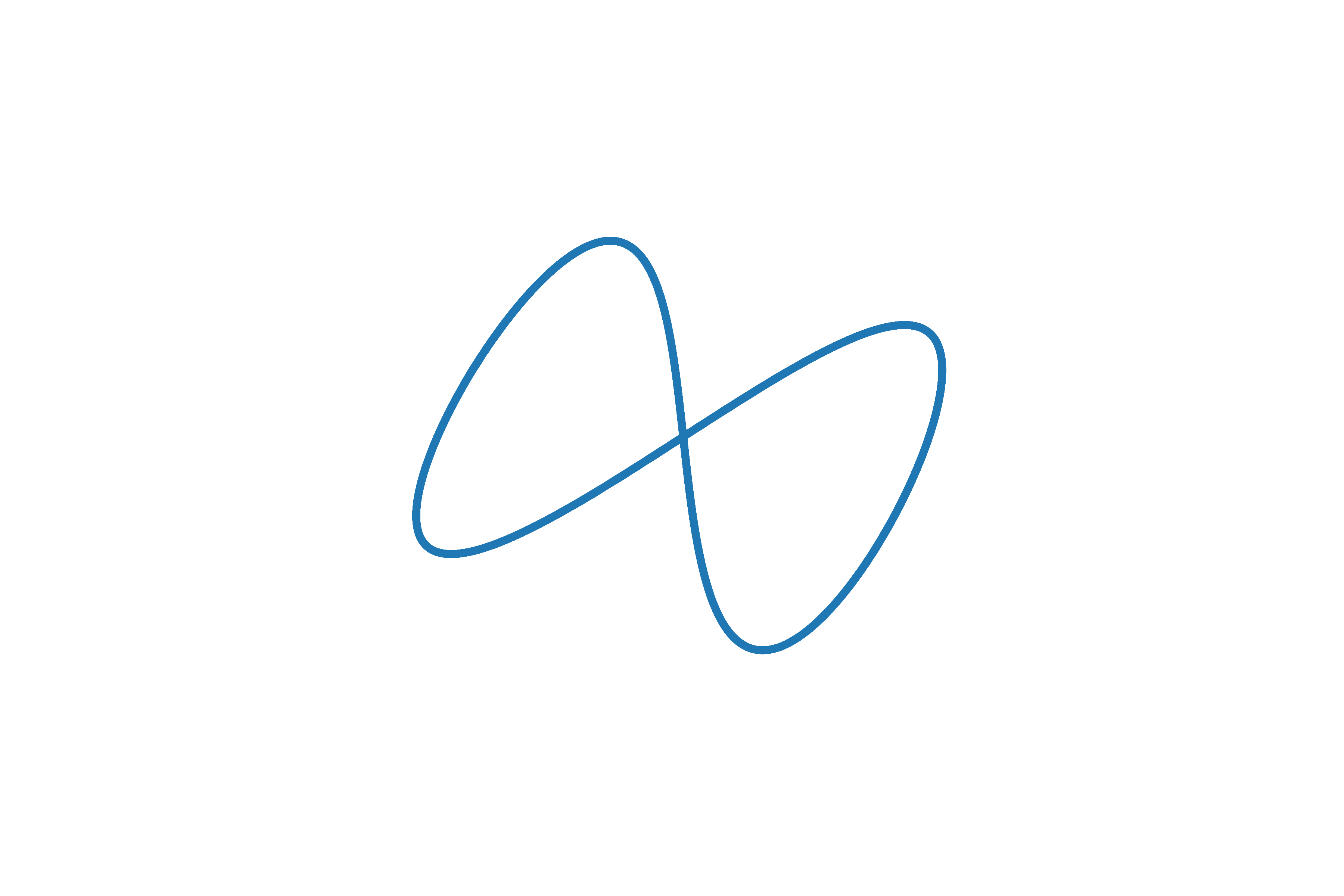}}} 
		& Nominal-MPC & 0.0583 & 0.0677 & 0.0643 \\
		& LR-MPC      & 0.0499 & 0.0567 & 0.0341 \\
		& DEKC-MPC    & \textbf{0.0410} & \textbf{0.0520} & \textbf{0.0237} \\ 
		\midrule		
		
		\multirow{3}{*}{\shortstack{Lissajous Curve \\ \includegraphics[width=1.5cm]{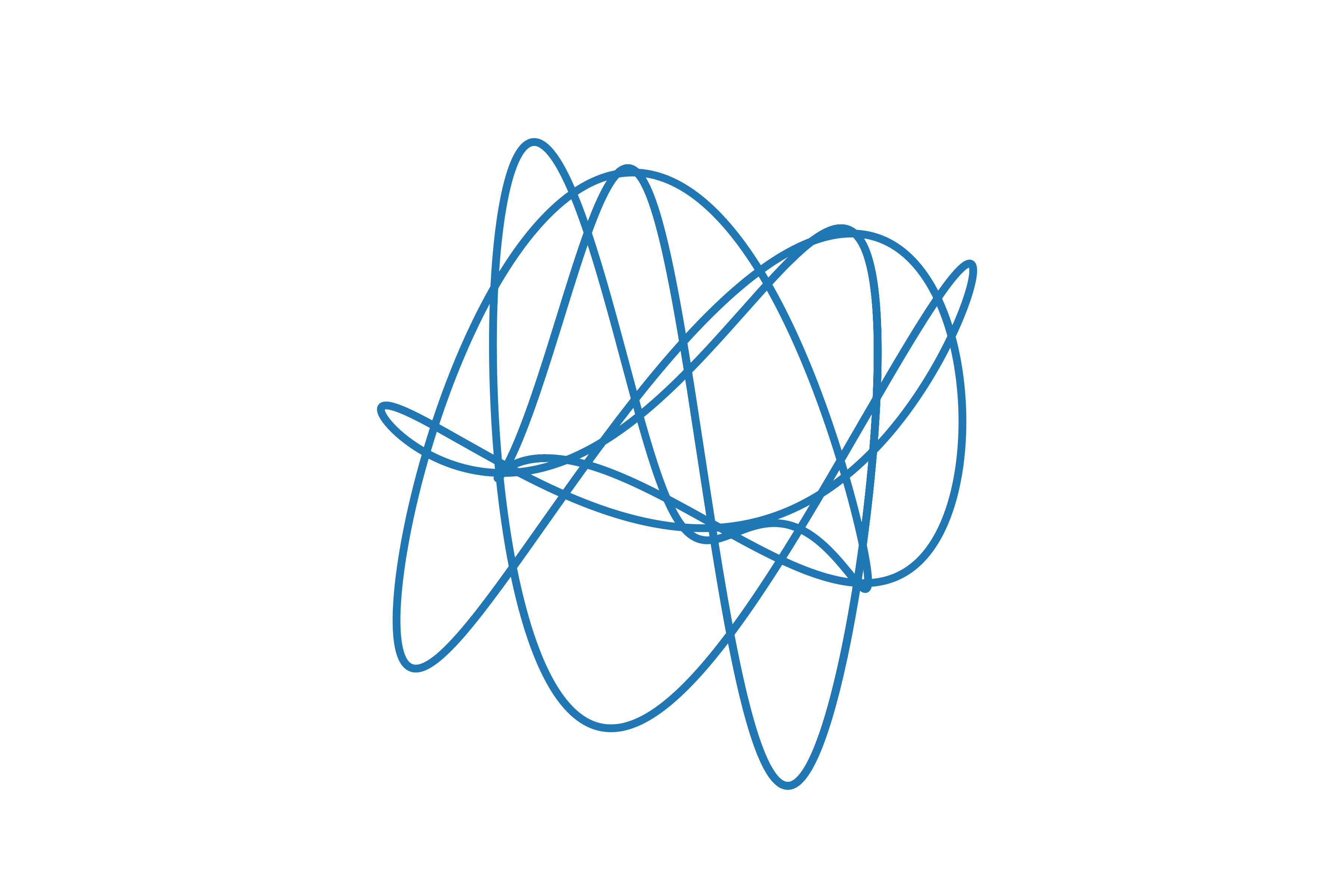}}} 
		& Nominal-MPC & 0.0745 & 0.0850 & 0.1238 \\
		& LR-MPC      & 0.0584 & 0.0625 & 0.0906 \\
		& DEKC-MPC    & \textbf{0.0319} & \textbf{0.0529} & \textbf{0.0619} \\ 
		\midrule
		
		\multirow{3}{*}{\shortstack{Trefoil Knot \\ \includegraphics[width=1.5cm]{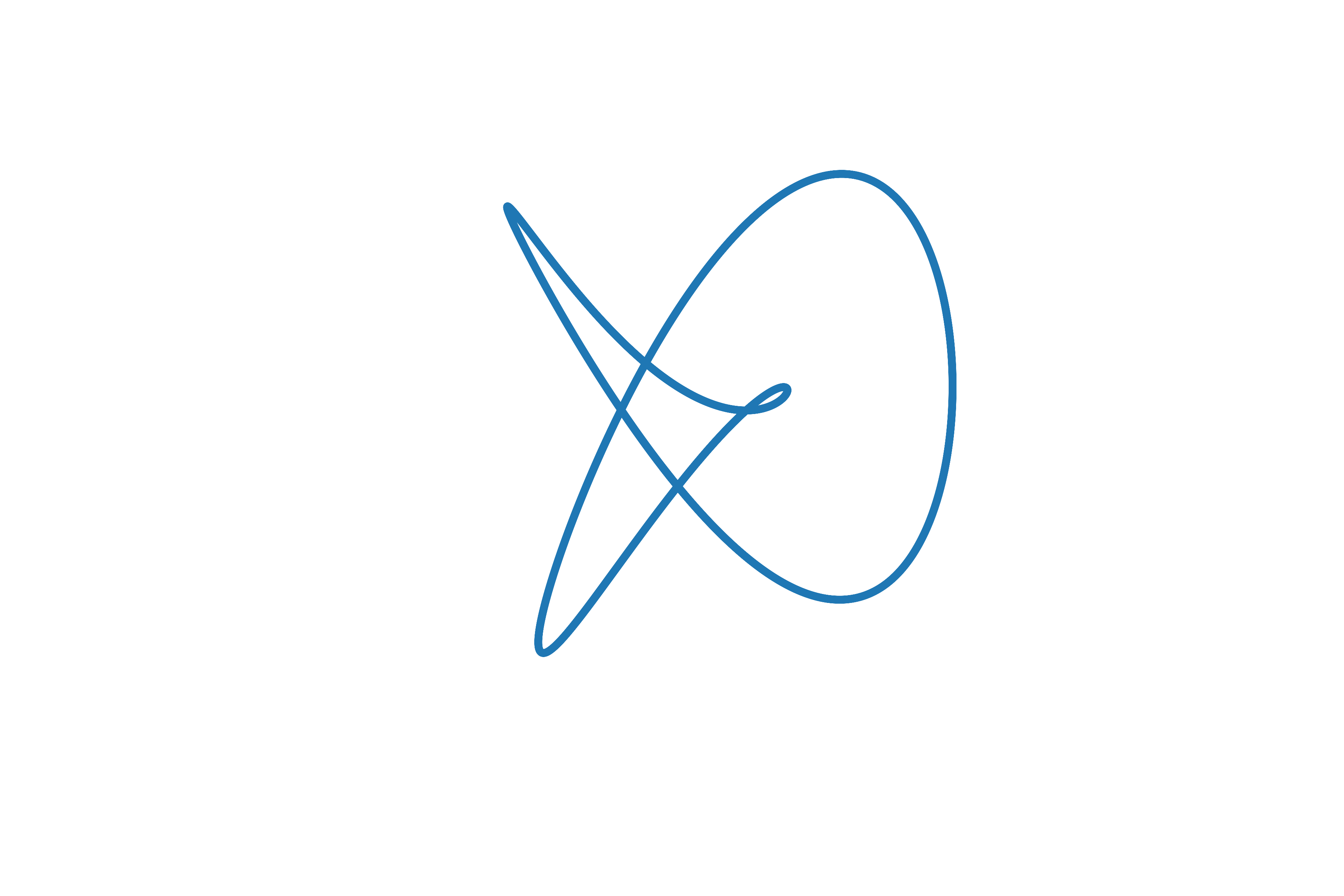}}} 
		& Nominal-MPC & 0.0609 & 0.0664 & 0.0707 \\
		& LR-MPC      & 0.0522 & 0.0509 & 0.0654 \\
		& DEKC-MPC    & \textbf{0.0502} & \textbf{0.0481} & \textbf{0.0486} \\ 
		\bottomrule
	\end{tabular}
\end{table}

\subsection{Results}
We first evaluate the tracking performance of three methods when the quadrotor is without any external payload. The objective is to verify whether the DEKC module can effectively capture the complex aerodynamic drag and residual dynamics during high-speed maneuvers. We tested a diverse set of aggressive 3D trajectories, including 3D Loop, Conical Spiral, Lemniscate, Lissajous Curve, and Trefoil Knot\footnote{The detailed description for tracking trajectory is given in Appendix  \ref{appendix_experiment_trajectory}.}. The quantitative tracking performance for the five agile trajectories is summarized in Table \ref{table_experiment_scene_a}. We report the Root Mean Square Error (RMSE) for the position tracking along the $x$, $y$, and $z$-axes ($E_x, E_y, E_z$). As shown in Table \ref{table_experiment_scene_a}, the proposed DEKC-MPC algorithm consistently achieves the lowest tracking errors across all five trajectories compared to the baseline methods. The Nominal-MPC exhibits the largest errors in almost all cases, particularly in the $z$-axis (e.g., $0.1188$ m for the 3D Loop and $0.1238$ m for the Lissajous Curve). This degradation is attributed to the controller's inability to account for significant aerodynamic drag and unmodeled dynamics during high-speed maneuvers. While the LR-MPC provides a noticeable improvement over the nominal baseline by compensating for linear aerodynamic drag, it still falls short of the performance achieved by DEKC-MPC. For instance, in the Conical Spiral trajectory, LR-MPC reduces the $z$-axis error to $0.0567$ m, but DEKC-MPC further reduces it to $0.0357$ m. This performance gap highlights the limitation of using fixed, hand-crafted basis functions. The LR-MPC struggles to capture the highly coupled, nonlinear residual dynamics that arise during agile maneuvers, whereas the DEKC can still successfully extracts these latent features. 

\begin{figure*}[htbp]
	\centering
	\subfigure[3D Loop]{\includegraphics[width=10.5pc]{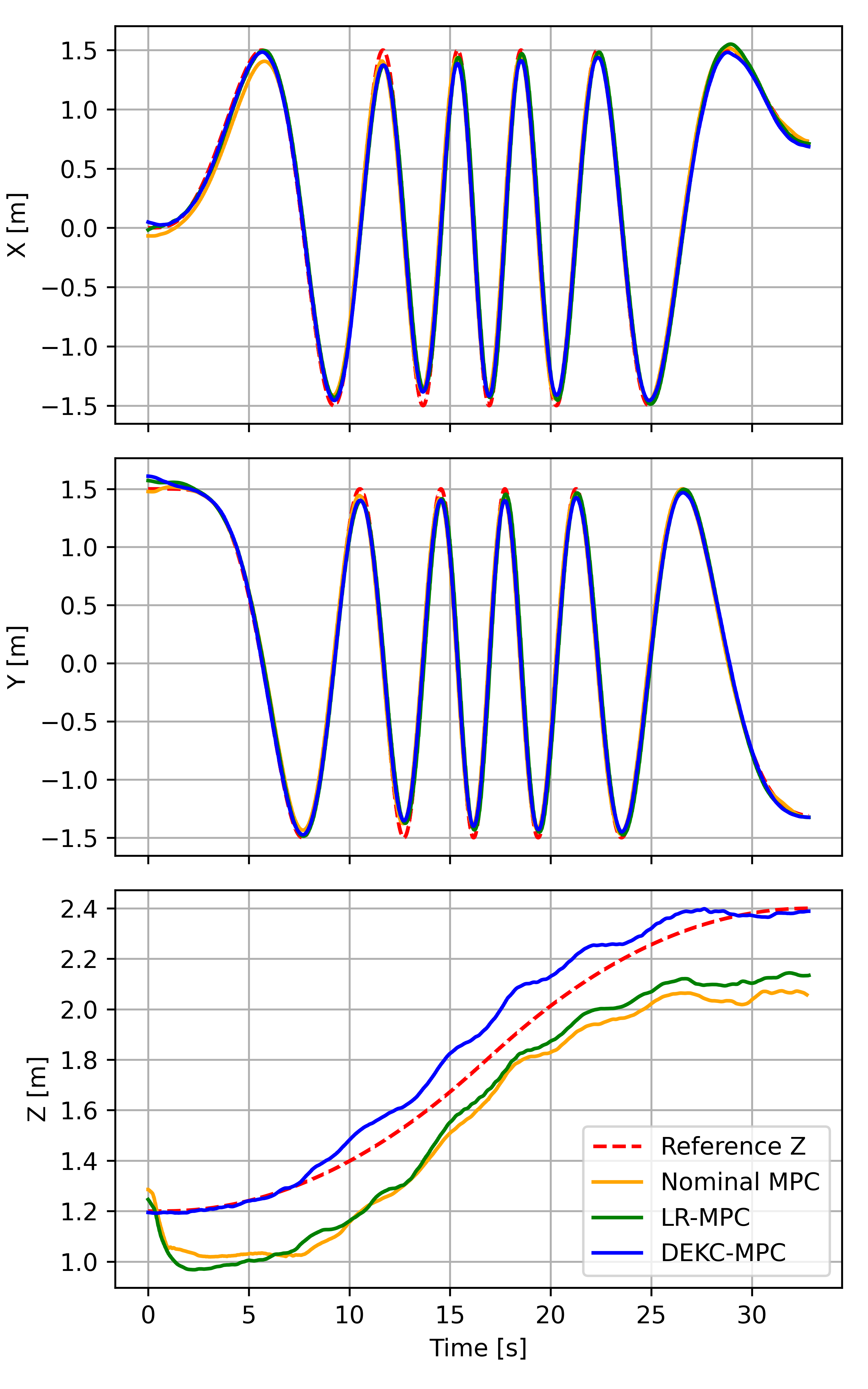}}
	\subfigure[Conical Spiral]{\includegraphics[width=10.5pc]{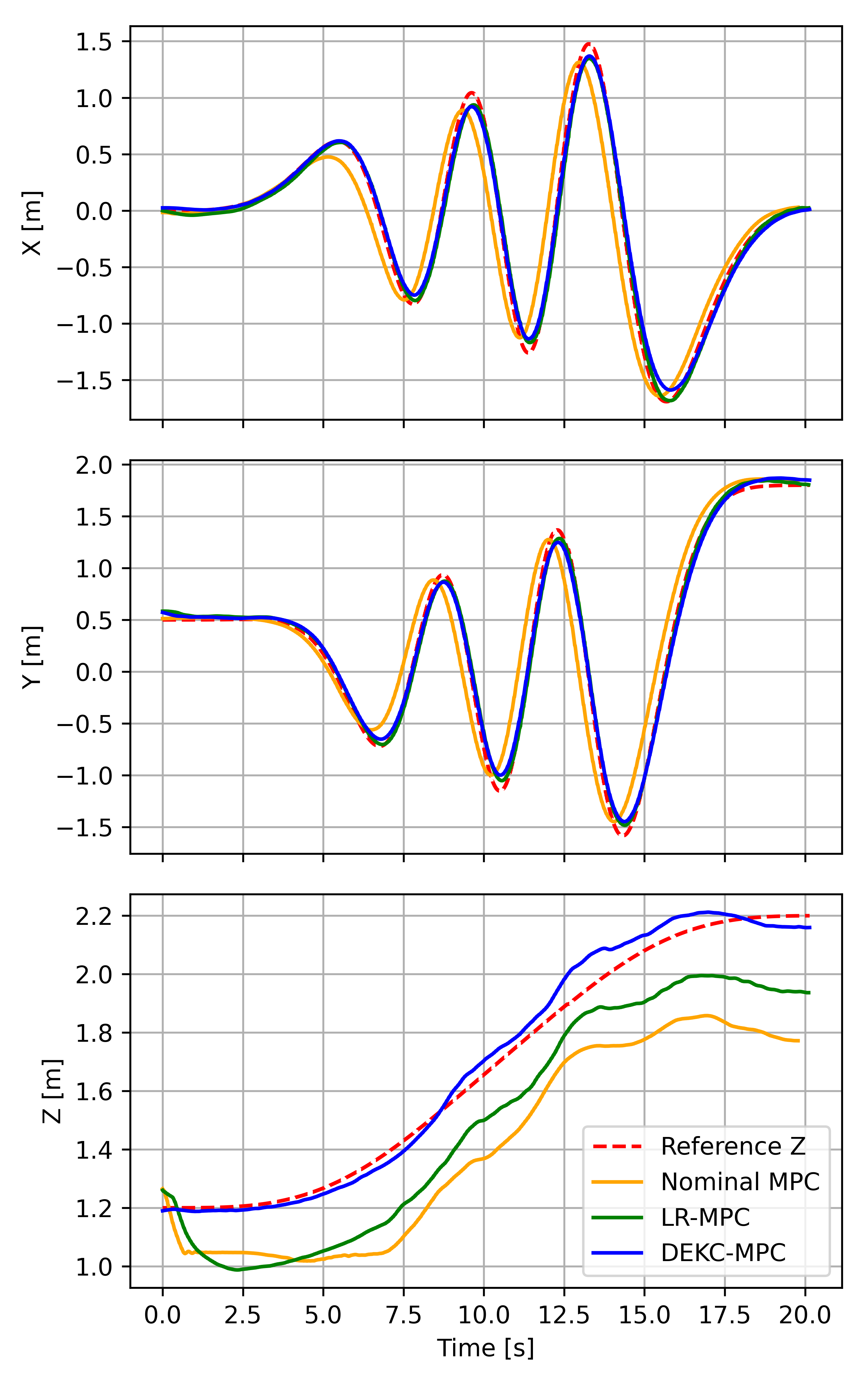}}
	\subfigure[Lemniscate]{\includegraphics[width=10.5pc]{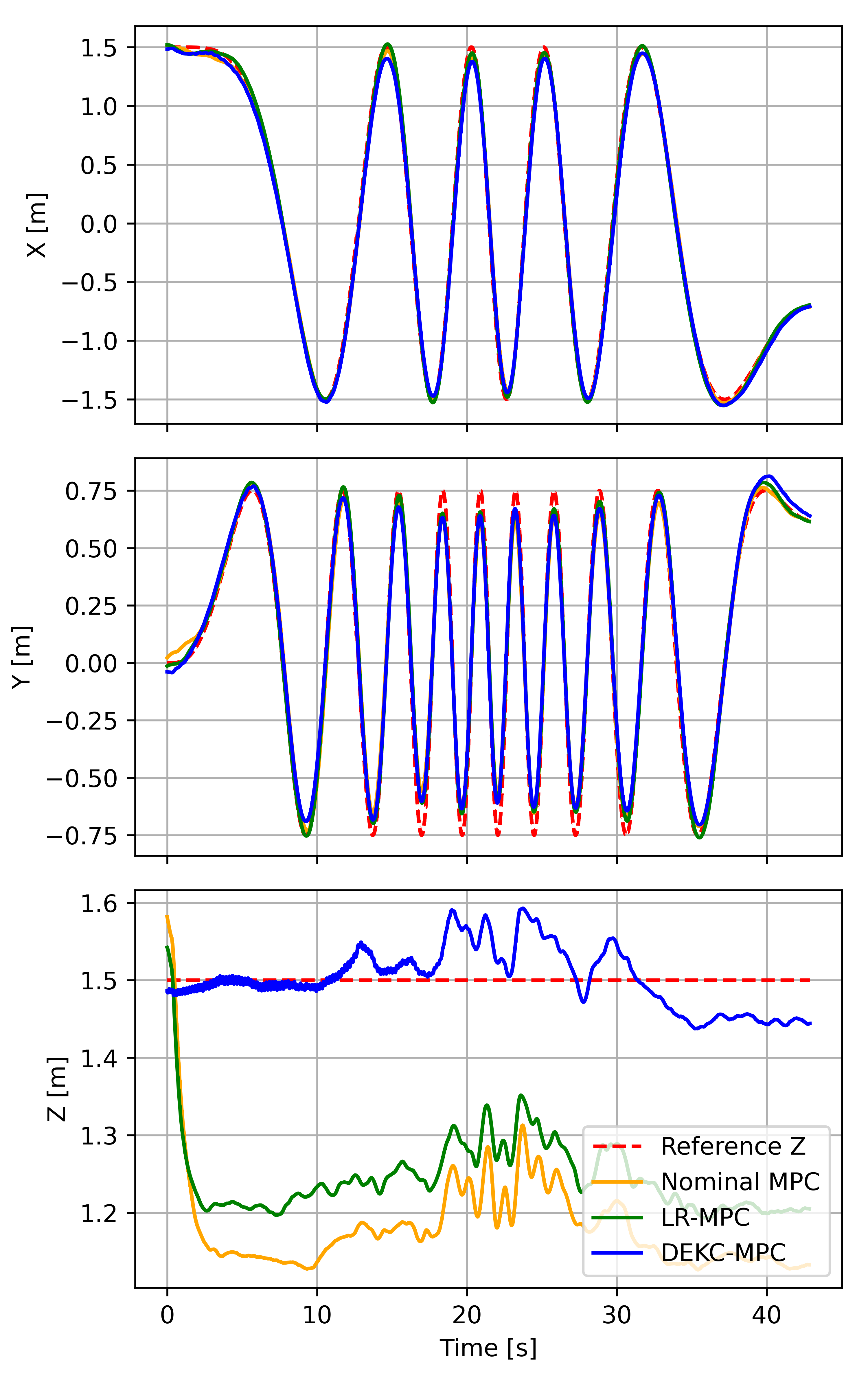}}
	\subfigure[Lissajous Curve]{\includegraphics[width=10.5pc]{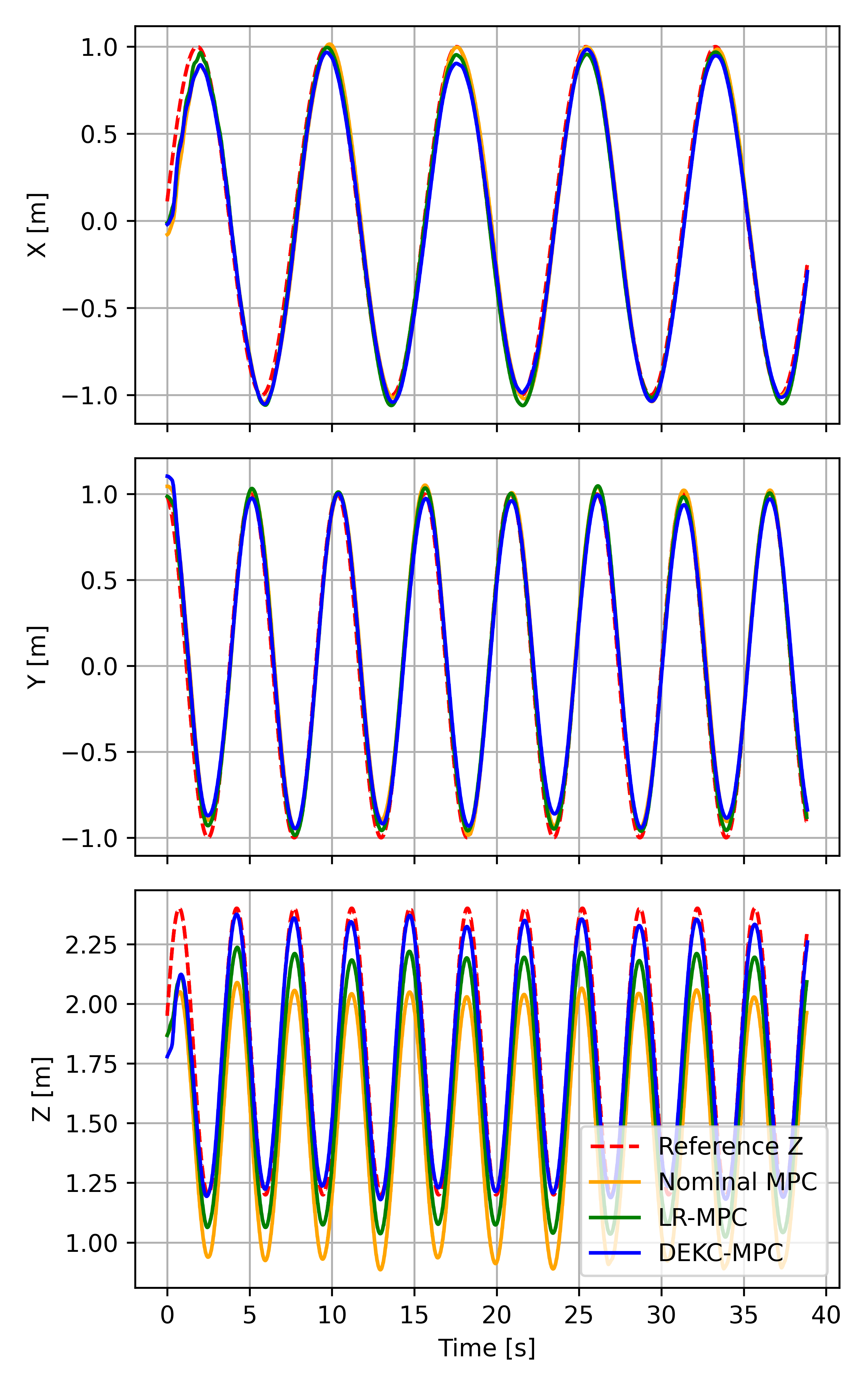}}
	
	\caption{The tracking results under the suspended load scenario. The red dashed lines represent the ground truth of reference trajectories. The solid lines depict the actual flight paths controlled by Nominal-MPC (Orange), LR-MPC (Green), and the proposed DEKC-MPC (Blue).}\label{fig_experiment_scene_b}
\end{figure*}

In the second scenario, we introduce significant unmodeled dynamics and parametric uncertainty by suspending a payload of 260 g from the quadrotor. This setup shifts the center of gravity and introduces pendulum-like oscillations that are not accounted for in the nominal model. The quadrotor was tasked to track the 3D Loop, Conical Spiral, Lemniscate, and Lissajous trajectories. The detailed tracking results are shown in Fig. \ref{fig_experiment_scene_b} and Table \ref{table_experiment_scene_a_payload}. The Nominal-MPC exhibits severe performance degradation. For instance, in the 3D Loop trajectory, the unmodeled gravitational force of the payload result in a substantial steady-state error, particularly in the vertical axis ($z$-axis). The LR-MPC offers marginal improvement but fails to consistently suppress the oscillations caused by the load swing, as the linear regression model struggles to capture the highly nonlinear coupling between the cable angle and the acceleration of quadrotor. The proposed DEKC-MPC maintains high-precision tracking across all four trajectories. This indicates that the DEKC framework successfully identifies the "equivalent disturbance" generated by the suspended load without any prior knowledge of the payload mass or cable length.

To understand the principle behind the performance gap for suspended load scenario, we analyze the system behavior in two distinct dynamic regimes. For the continuous centrifugal maneuvers (3D Loop and Conical Spiral), the suspended load swings outward due to centrifugal force, creating a persistent bias force on the quadrotor. The Nominal-MPC, unaware of this force, drifts radially outward and loses altitude. The DEKC estimator, however, rapidly converges to this quasi-static disturbance component. It effectively "learns" the shifted center of mass and the additional thrust required to counteract the downward pull of the load, ensuring the drone adheres tightly to the reference path. For the agile maneuvers (Lemniscate and Lissajous), they involve rapid changes in velocity and direction. This is the most challenging condition for the LR-MPC, since the dynamics of disturbance change faster than the sliding window can adapt. In contrast, the DEKC-MPC leverages the ability of Koopman operator to predict the evolution of the disturbance. Based on the learned disturbance dynamics, the MPC generates control actions that actively dampen the swing. This active compensation is evident in the Lissajous experiment, where the DEKC-MPC significantly reduces the oscillatory overshoot at the turning points compared to the other two baseline methods.

\begin{figure}[!t]
	\centering
	{\includegraphics[width=20pc]{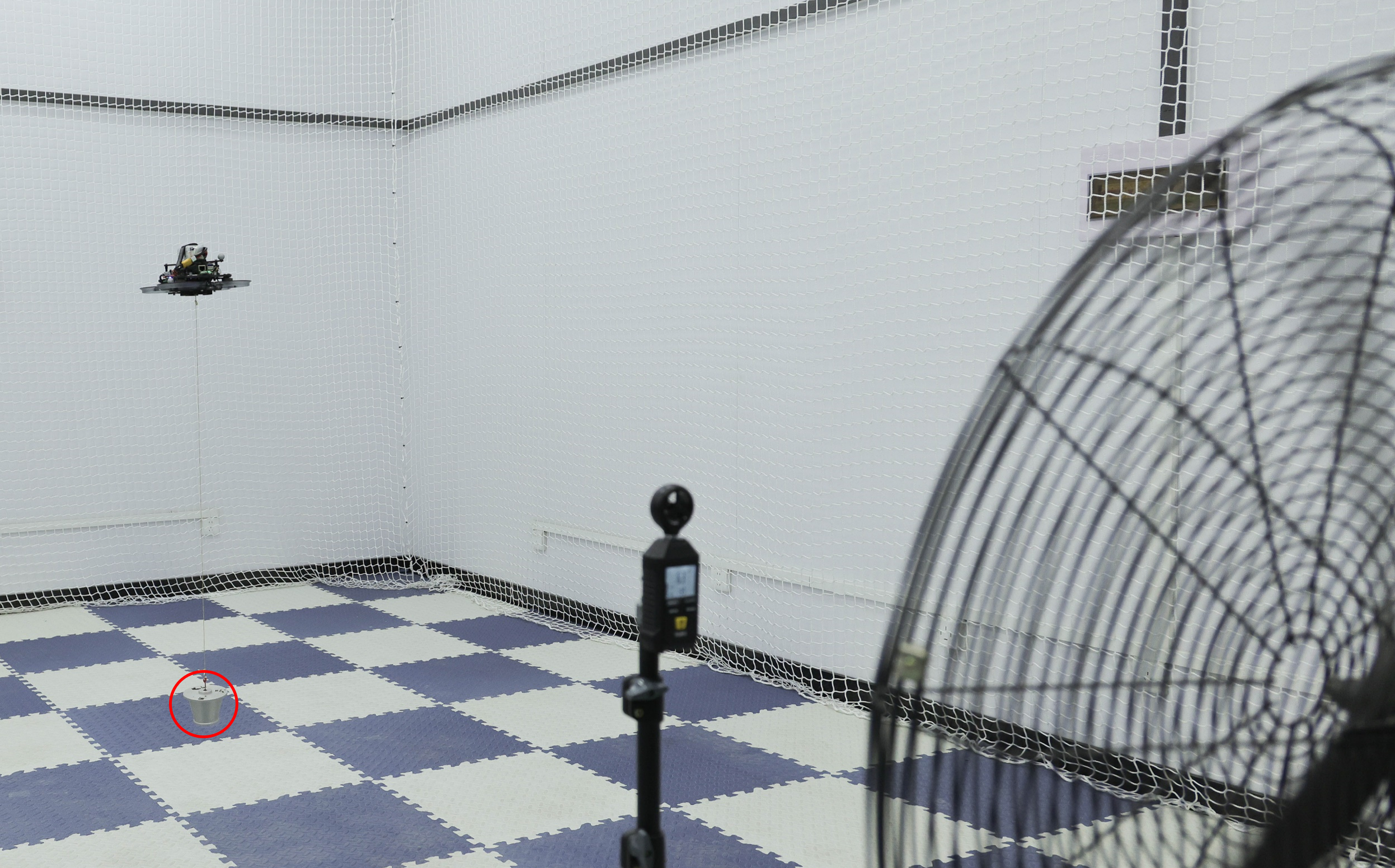}}
	\caption{Experimental setup for the compound disturbance scenario. The quadrotor is tasked with hovering while carrying a suspended payload (red object) and facing a turbulent wind field generated by an industrial fan (right).}
	\label{hovering_experiment}
\end{figure}

\begin{figure*}[!t]
	\centering
	{\includegraphics[height=11pc]{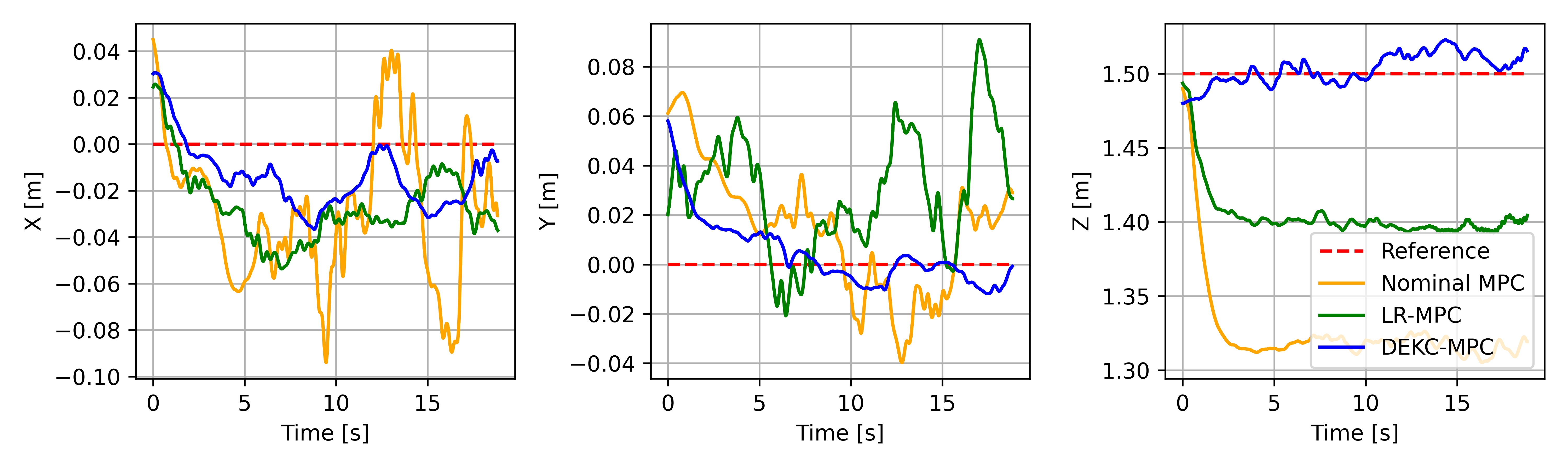}}
	\caption{Position tracking results under compound disturbances. The quadrotor performs a hovering task while subject to both internal parametric uncertainty (suspended load) and external environmental forces (wind gusts). The curves depict the position deviation from the setpoint for Nominal-MPC (Orange dashed line), LR-MPC (Green dotted line), and DEKC-MPC (Blue solid line). }
	\label{fig_experiment_scence_c}
\end{figure*}

The final and most challenging scenario involves compound disturbances. The quadrotor is required to maintain a stable hover while carrying the same suspended payload in the second scenario and being subjected to external wind gusts. We positioned a high-power industrial fan at a distance of approximately 2.0 meters from the target hovering point. The fan generates an intermittent and turbulent wind field with speeds reaching up to 6.0 m/s. The configuration of the experiment is illustrated in Fig. \ref{hovering_experiment}. The hovering results are shown in  Fig. \ref{fig_experiment_scence_c} and the last row of Table \ref{table_experiment_scene_a_payload}. The Nominal-MPC fails to maintain a safe hover. The unmodeled swing of the payload, coupled with the wind gusts, pushes the drone away from the setpoint. The LR-MPC shows improved stability but suffers from a sluggish response. Due to the reliance on a sliding window for regression, the linear estimator lags behind the rapidly changing aerodynamic forces, leading to a persistent steady-state error. In contrast, the proposed DEKC-MPC demonstrates superior disturbance rejection capability. The proposed algorithm rapidly identifies the "total equivalent disturbance" (combining external wind gusts and cable forces) and predicts its evolution. As a result, the controller takes preemptive actions to counteract the disturbances before they significantly displace the vehicle. Specifically, the most critical improvement is observed in the vertical direction ($z$-axis), where the unmodeled payload weight causes the Nominal MPC and LR-MPC to suffer from large steady-state errors of 0.1769 m and 0.0976 m, respectively. In contrast, the DEKC-MPC achieves an RMSE of only 0.0107 m, effectively eliminating the height drop by compensating for the unknown gravitational pull. Similarly, in the horizontal plane ($x$ and $y$ axes) dominated by wind disturbances, the DEKC-MPC maintains tight tracking precision with RMSEs below 0.02 m (0.0198 m and 0.0138 m), significantly outperforming the baselines which exhibit drifts of up to 0.0422 m. These results validate that the DEKC framework successfully learns to decouple and suppress the compound effects of wind gusts and payload dynamics.

\begin{table}[!t]
	\centering 
	\renewcommand{\arraystretch}{1.4}
	\caption{RMSE Results of Tracking Performance with Payload} \label{table_experiment_scene_a_payload}
	\begin{tabular}{lcccc} 
		\toprule
		\textbf{Trajectory}  & \textbf{Method} & \textbf{$E_x$} [m]  & \textbf{$E_y$} [m] & \textbf{$E_z$} [m] \\ 
		\midrule
		
		\multirow{3}{*}{\shortstack{3D Loop}}  
		& Nominal-MPC &  0.0993 &0.0855 & 0.2346      \\
		& LR-MPC & 0.0793 & 0.0697  & 0.2037   \\
		& DEKC-MPC &\textbf{ 0.0550} & \textbf{0.0553} & \textbf{0.0522}    \\
		\midrule
		
		\multirow{3}{*}{\shortstack{Conical Spiral}} 
		& Nominal-MPC & 0.0969 & 0.0939 & 0.2908 \\
		& LR-MPC      & 0.0734 & 0.0707 & 0.1882 \\
		& DEKC-MPC    & \textbf{0.0633} & \textbf{0.0522} & \textbf{0.0448} \\ 
		\midrule	
		
		\multirow{3}{*}{\shortstack{Lemniscate}} 
		& Nominal-MPC & 0.0925 & 0.0786 & 0.3242 \\
		& LR-MPC      & 0.0695 & 0.0655 & 0.2591 \\
		& DEKC-MPC    & \textbf{0.0543} & \textbf{0.0622} & \textbf{0.0393} \\ 
		\midrule		
		
		\multirow{3}{*}{\shortstack{Lissajous Curve}} 
		& Nominal-MPC & 0.0981 & 0.0971 & 0.3212 \\
		& LR-MPC      & 0.0707 & 0.0697 & 0.1779 \\
		& DEKC-MPC    & \textbf{0.0646} & \textbf{0.0611} & \textbf{0.0689} \\ 
		\midrule
		
		\multirow{3}{*}{\shortstack{Hovering (Wind)}} 
		& Nominal-MPC &  0.0422 &0.0287 & 0.1769      \\
		& LR-MPC & 0.0312 & 0.0382  & 0.0976   \\
		& DEKC-MPC &\textbf{ 0.0298} & \textbf{0.0138} & \textbf{0.0107}    \\
		\bottomrule
	\end{tabular}
\end{table}

\section{Conclusion}\label{section_conclusion}
This paper presented a differentiable data-enabled Koopman framework to address the challenge of high-precision control under unstructured uncertainties. The effectiveness of DEKC were rigorously validated on two robotic platforms. Numerical simulations on a TSR demonstrated the superior capability in handling highly coupled nonlinearities. Furthermore, real-world experiments on a quadrotor substantiated the practical robustness of the method. The results confirmed that DEKC significantly outperforms conventional baselines in tracking agile trajectories subject to compound disturbances induced by aerodynamics and suspended payload. We achieve this by treating the unknown uncertainties not as static errors, but as dynamical systems with learnable temporal evolution. Through a deep neural network-parameterized lifting mapping and an online backward gradient update mechanism, the framework successfully linearizes complex residual dynamics, enabling predictive compensation within the MPC or feedback control structure. Our future works will focus on extending this framework to distributed formation control of multiple agents, investigating how learned uncertainty dynamics can be shared among robots to enhance collective adaptation.

\section*{Appendix}

\begin{table*}[h] 
	\centering
	\caption{The Kinematic Constraints of Reference Trajectories}
	\label{tab_traj_specs}
	\renewcommand{\arraystretch}{1.3} 
	\begin{tabular}{lcccc}
		\toprule
		\textbf{Trajectory} & \textbf{Spatial Bounds} $\bm{P}$ [m] & \textbf{Max. Velocity} $\bm{v}_{\max}$ [m/s] & \textbf{Max. Accel.} $\bm{a}_{\max}$ [m/s$^2$] & \textbf{Duration} [s] \\
		& $(x, y, z)$ & $(v_x, v_y, v_z)$ & $(a_x, a_y, a_z)$ & $T$ \\
		\midrule
		
		3D Loop & 
		$[-1.50, 1.50] \times [-1.50, 1.50] \times [1.20, 2.40]$ & 
		$3.05, \, 3.05, \, 0.07$ & 
		$6.22, \, 6.09, \, 0$ & 
		$34$ \\
		
		Conical Spiral & 
		$[-1.69, 1.48] \times [-1.58, 1.80] \times [1.20, 2.20]$ & 
		$2.36, \, 2.36, \, 0.09$ & 
		$3.96, \, 4.06, \, 0.02$ & 
		$22$ \\
		
		Lemniscate & 
		$[-1.50, 1.50] \times [-0.75, 0.75] \times [1.50, 1.50]$ & 
		$2.02, \, 2.02, \, 0$ & 
		$2.69, \, 5.47, \, 0$ & 
		$44$ \\
		
		Lissajous Curve & 
		$[-1.00, 1.00] \times [-1.00, 1.00] \times [1.20, 2.40]$ & 
		$0.80, \, 1.20 \, 1.08$ & 
		$0.64, \, 1.44, \, 1.94$ & 
		$40$ \\
		
		Trefoil Knot & 
		$[-1.37, 1.37] \times [-1.50, 1.03] \times [1.00, 3.00]$ & 
		$1.77, \, 1.61, \, 2.13$ & 
		$2.18, \, 1.97, \, 4.52$ & 
		$35$ \\
		
		\bottomrule
	\end{tabular}
\end{table*}

\subsection{Formulations of Baseline Disturbance Observers in Section \ref{section_simulation_A}}
For completeness, this appendix summarizes the mathematical formulations of the three baseline disturbance observers used for comparison in the numerical simulations: the adaptive disturbance observer (ADO), the nonlinear disturbance observer (NDO), and the sliding-mode disturbance observer (SMDO). All observers are implemented using their standard forms reported in the literature, and their parameters are listed in Table \ref{baseline_do_params}. 

The adaptive disturbance observer \cite{Wang2017a} is given by
\begin{align}
	\begin{cases}
		&\dot{\bm{z}} = -a\tilde{\bm{x}}+\bm{f}(\bm{x})+\bm{g}_u(\bm{x})u+\bm{g}_d(\bm{x})\hat{\bm{d}}
		\\
		&\bm{g}_d(\bm{x})\hat{\bm{d}}=-\varUpsilon(T_s)\tilde{\bm{x}}(kT_s)\\
		&\varUpsilon(t)=\frac{a}{e^{aT_s}-1}
	\end{cases}
\end{align}
where $\tilde{\bm{x}}=\bm{z}-\bm{x}$, $t\in [kT_s,(k+1)T_s)$, $a$ is the adaption gain, and $T_s$ is the sampling period.

The nonlinear disturbance observer \cite{Li2014a} is given by
\begin{align}
	\begin{cases}
		&\dot{\bm{z}} = -\bm{L}\bm{g}_d(\bm{x})\bm{z}-\bm{L}[\bm{g}_d(\bm{x})\bm{L}\bm{x}+\bm{f}(\bm{x})+\bm{g}_u(\bm{x})]\\
		&\bm{\hat{d}}=\bm{z}+\bm{L}\bm{x}
	\end{cases}
\end{align}
where $\bm{L}$ is the observer gain.

The sliding-mode disturbance observer \cite{Lu2012} is given by
\begin{align}
	\begin{cases}
		& \dot{\bm{z}}_0=\bm{v}_0+\bm{f}(\bm{x})+\bm{g}_u(\bm{x})\bm{u} \\
		& \bm{v}_0 = -\lambda_0 \bm{L}^{1/3} |\bm{z}_0-\bm{x}|^{2/3}\cdot sign(\bm{z}_0-\bm{x})+\bm{z}_1 \\
		& \dot{\bm{z}}_1=\bm{v}_1 \\
		& \dot{\bm{v}}_1 = \lambda_1 \bm{L}^{1/3}|\bm{z}_1-\bm{v}_0|^{2/3}\cdot sign(\bm{z}_1-\bm{v}_0)+\bm{z}_2 \\
		& \dot{\bm{z}}_2 = -\lambda_2 \bm{L}\cdot  sign(\bm{z}_2-\bm{v}_1)	
	\end{cases}
\end{align}
where $\bm{L}$ is the sliding gain and $\bm{\lambda}=[\lambda_0\ \lambda_1\ \lambda_2]$ is the observer gain.

\subsection{Tracking Trajectory for Real-World Experiments}\label{appendix_experiment_trajectory}
The kinematic constraints, including the bounds of position, velocity, and acceleration, are summarized in Table \ref{tab_traj_specs} to demonstrate the aggressiveness of the flight maneuvers.

\bibliographystyle{IEEEtran}
\bibliography{bibtex.bib}

\end{document}